\documentclass[aps,preprint,showpacs,floats,epsf,epsfig,nofootinbib,12pt]{revtex4}
\usepackage{graphicx}% Include figure files
\usepackage{epsfig}
\usepackage[dvips]{color}

\def\beq{\begin{eqnarray}}
\def\eeq{\end{eqnarray}}

\def\la{\langle}
\def\ra{\rangle}

\begin{document}

\title{Is $Z_c(3900)$  a molecular state}

\vspace{1cm}

\author{ Hong-Wei Ke$^{1}$   \footnote{khw020056@hotmail.com}, Zheng-Tao Wei$^2$ \footnote{weizt@nankai.edu.cn} and
        Xue-Qian Li$^2$\footnote{lixq@nankai.edu.cn}  }

\affiliation{  $^{1}$ School of Science, Tianjin University, Tianjin 300072, China \\
  $^{2}$ School of Physics, Nankai University, Tianjin 300071, China }

\vspace{12cm}

\begin{abstract}
Assuming the newly observed $Z_c(3900)$ to be a molecular state of
$D\bar D^*(D^{*} \bar D)$, we calculate the  partial widths of
$Z_c(3900)\to J/\psi+\pi;\; \psi'+\pi;\; \eta_c+\rho$ and $D\bar
D^*$ within the light front model (LFM). $Z_c(3900)\to J/\psi+\pi$
is the channel by which $Z_c(3900)$ was observed, our calculation
indicates that it is indeed one of the dominant modes whose width
can be in the range of a few MeV depending on the model
parameters. Similar to $Z_b$ and $Z_b'$, Voloshin suggested that
there should be a resonance $Z_c'$ at 4030 MeV which can be a
molecular state of $D^*\bar D^*$. Then we go on calculating its
decay rates to all the aforementioned final states and as well the
$D^*\bar D^*$. It is found that if $Z_c(3900)$ is a molecular
state of ${1\over\sqrt 2}(D\bar D^*+D^*\bar D)$, the partial width
of $Z_c(3900)\to D\bar D^*$ is rather small, but the rate of
$Z_c(3900)\to\psi(2s)\pi$ is even larger than $Z_c(3900)\to
J/\psi\pi$. The implications are discussed and it is indicated
that with the luminosity of BES and BELLE, the experiments may
finally determine if $Z_c(3900)$ is a molecular state or a
tetraquark.

\pacs{14.40.Lb, 12.39.Mk, 12.40.-y}

\end{abstract}

\maketitle

\section{Introduction}
Recently the BES collaboration\cite{Ablikim:2013mio} has claimed
that a new resonance is observed in the invariance mass spectrum
of $J/\psi\pi^{\pm}$  by studying the process $e^+e^-\to
J/\psi\pi^+\pi^-$ at $\sqrt s=4.26$ GeV, which is referred as
$Z_c(3900)$ with its mass and width being $(3.899\pm3.6\pm4.9)$
GeV and $(46\pm10\pm10)$ MeV respectively.  The
Belle\cite{Liu:2013dau} and CLEO\cite{Xiao:2013iha} also reported
the same new structure.  Since the resonance is charged it cannot
be a charmonium, but its mass and decay modes imply that it has a
hidden charm-anticharm structure, therefore it must be  an exotic
state. In fact, before this discovery, two bottomonium-like
charged resonances $Z_b(10610)$ and $Z_b(10650)$ were observed by
BELLE\cite{Choi:2003ue} and confirmed by BABAR
\cite{Collaboration:2011gj}. Because they cannot be bottomonia, a
reasonable postulate is that they are exotic states with
constituents of $b\bar b u\bar d (d\bar u)$, e.g. they may be
molecular states or tetraquarks etc. Observation of similar
charged meson $Z_c(3900)$ indicates that at the charm energy range
there exist similar exotic states. It intrigues enormous interests
of theorists
\cite{Chen:2013coa,Cui:2013yva,Zhang:2013aoa,Wang:2013cya,Wilbring:2013cha,Voloshin:2013dpa}.
For $Z_c(3900)$ some authors suggest it to be a molecular
state\cite{Cui:2013yva,Zhang:2013aoa,Wang:2013cya,Wilbring:2013cha},
whereas some others think it as tetraquark or a mixture of the two
states\cite{Voloshin:2013dpa}. Which one is the true
configuration? The answer can only be obtained from experimental
measurements. Namely different structures would result in
different decay rates for various channels. Therefore, by assuming
a special structure, we predict its decay rates for those possible
modes, then the theoretical predictions will be tested by further
more accurate measurements and their consistency with data would
tell us if the postulation about the hadron structure is
reasonable. In this paper we will study the strong decays of
$Z_c(3900)$ which is assumed to be a molecular state with the
quantum number $I^G(J^P)=1^+(1^+)$. Assuming $Z_c(3900)$ to have
constituents of $D\bar D^*(D^*\bar D)$,  we investigate the decays
$Z_c(3900)\rightarrow J/\psi\pi;\;\psi'\pi;\;\eta_c\rho$ and
$D\bar D^*$ under this assignment.

Comparing with $Z_b(10610)$ and $Z_b(10650)$, $Z_c(3900)$ has the
same light degrees of freedom, so that one may expect another
resonance to exist around $4030$
MeV \cite{Voloshin:2013dpa} and it could be of the molecular
structure of $D^*\bar D^*$. With that assignment we calculate the
decay rates of $Z_c(4030)$ via the aforementioned decay modes for
$Z_c(3900)$ as well as $D^*\bar D^*$ because this channel is open at the energy 4030 MeV
within the same theoretical framework.

In this work, we will extend the light front quark
model (LFQM) which was thoroughly studied in literature
\cite{Jaus,Ji:1992yf,Cheng:2004cc,Cheng:1996if,Cheng:2003sm,Choi:2007se,
Hwang:2006cua,Ke:2007tg,Ke:2009ed,Li:2010bb,Ke:2013zs}
to investigate the decays of a molecular state. Initially the
authors\cite{Jaus,Ji:1992yf} constructed the light front quark
model (LFQM)  which is used to study processes where only mesons are involved. Later Cheng et. al.
extended the framework to explore the decays of pentaquark\cite{Cheng:2004cc}.
Along the line we have extended the model to calculate the decay rates of baryons \cite{Ke:2007tg}.
With the LFQM, the theoretical predictions are reasonably consistent with data, it implies
that the applications of LFQM to various situations at charm and bottom energy regions are
comparatively successful. This success inspires us to extend the light front model
to study decays of molecular states.

In this approach the constituents are two mesons instead of a
quark and an antiquark  in the light front frame. In the
covariant case the constituents are not on-shell. The effective
interactions between the two concerned constituent mesons
are that often adopted when one studies the effects
of final state
interactions\cite{Haglin:1999xs,Oh:2000qr,Lin:1999ad,Deandrea:2003pv,Meng:2007cx,Yuan:2012zw}.
Namely, by studying such processes, one can extract the effective
coupling constants from the data. Since for the molecular states
the constituents and interactions are different from the case for
quarks, we need to modify the LFQM and then apply the new version
to study the exotic hadrons. In this paper we will deduce the form
factors for the two-body decays of a molecular state of
$I^G(J^P)=1^+(1^+)$, and use them to estimate the decay widths of
$Z_c(3900)\rightarrow J/\psi\pi;\; \psi'\pi;\; \eta_c\rho$ and $
D\bar D^*$ by assuming $Z_c(3900)$ to be a molecular state of $D\bar D^*$,
then we calculate the rates of $Z_c(4030)\rightarrow J/\psi\pi;\;
\psi'\pi;\; \eta_c\rho$ and $D\bar D^*;\; D^*\bar D^*$.

In our calculation, we  keep the  $q^+=0$ condition i.e. $q^2<0$ where
one of the final mesons ($\pi$ or $D$) is off-shell, thus the obtained
form factors are space-like, i.e. unphysical. Then an analytical extension from the
space-like region to the time-like region is applied. Letting the
meson be on-shell one can get the physical form factor and
calculate the corresponding decay widths. The numerical results will offer us information about the
structure of $Z_c(3900)$ and the possible $Z_c(4030)$.

After the introduction we derive the form factor for transitions
$Z_c(3900)\rightarrow J/\psi\pi;\; \psi'\pi;\; \eta_c\rho$ and $
D\bar D^*$ and $Z_c(4030)\rightarrow J/\psi\pi;\; \psi'\pi;\; \eta_c\rho$ and $D\bar D^*;\; D^*\bar D^*$  in section II. Then we
numerically evaluate the relevant form factors and decay widths
in Sec. III, where all input parameters are presented. At last we
discuss the implications of the numerical results possibilities, then
finally, we draw our conclusion even though it is not very definite so far. Some details About the adopted approach are collected in the appendix.

\section{the strong decays of $Z_c(3900)$ as a $1^{+}$ $D\bar D^*$ molecular state}
In this section we study the strong decays of a $1^{+}$ $D\bar
D^*$ molecular state in the light-front model. In
Ref.\cite{Jaus,Ji:1992yf,Cheng:2003sm} the model is used to
explore some meson decays. In this paper we extend it to study a
molecular state and the interactions between mesons are regarded
as effective ones. The configuration of $D\bar D^*$ molecular
state is $\frac{1}{\sqrt{2}}(D\bar D^*+\bar D D^*)$. By the
Feynman diagrams it is also noted that the topological structure
for $Z_c(3900)\to J/\psi \pi;\; \psi'\pi;\; \eta_c\rho$ is
different from that for $Z_c(3900)\to D\bar D^*$, so we deal with
them  separately.

\subsection{$Z_c(3900)\rightarrow J/\psi\pi\,(\rho\eta_c \,{\rm or} \,\psi'\pi)$}

\begin{figure}
\begin{center}
\begin{tabular}{ccc}
\scalebox{0.6}{\includegraphics{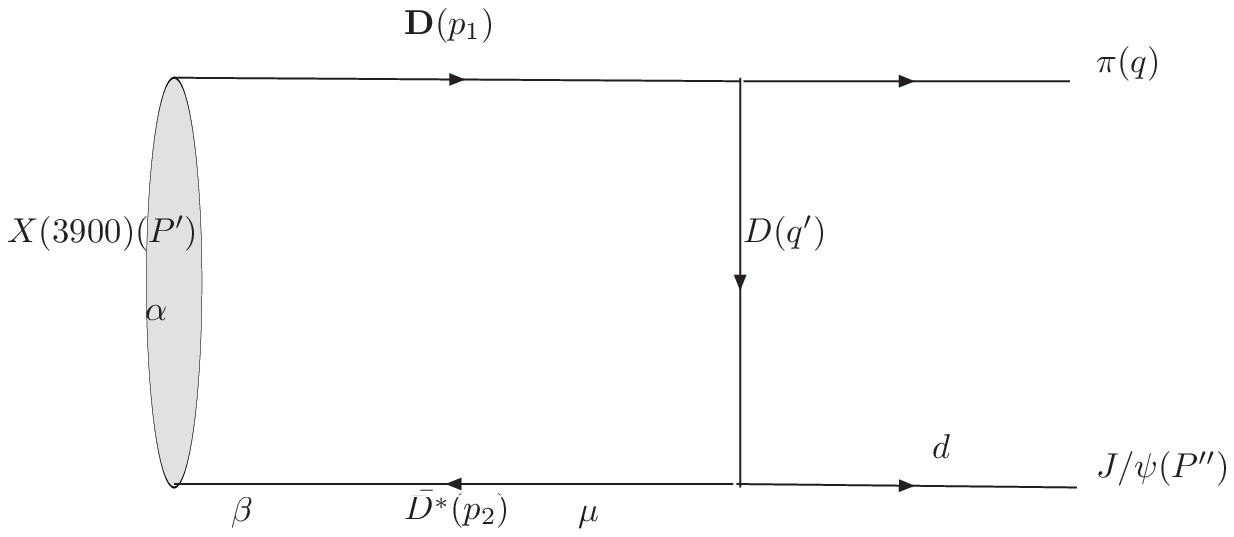}}\scalebox{0.6}{\includegraphics{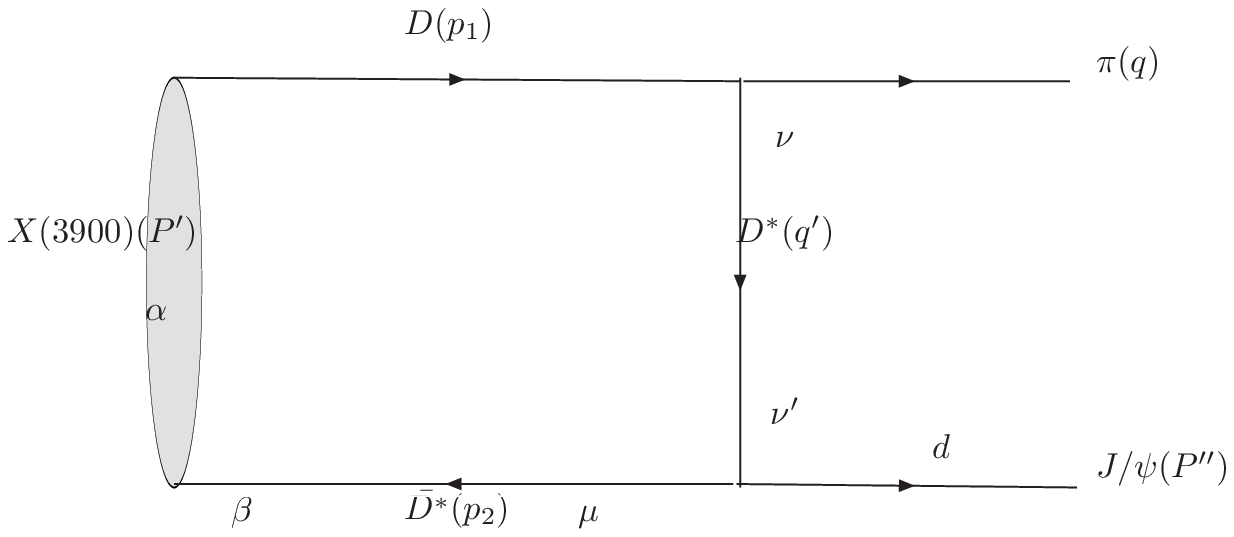}}\\
\qquad\qquad(a)\qquad\qquad\qquad\qquad\qquad\qquad\qquad
\,\,\,\,\,\,\,\,\,\,\,\,\,\,\,\,\,\,(b)\\
\scalebox{0.6}{\includegraphics{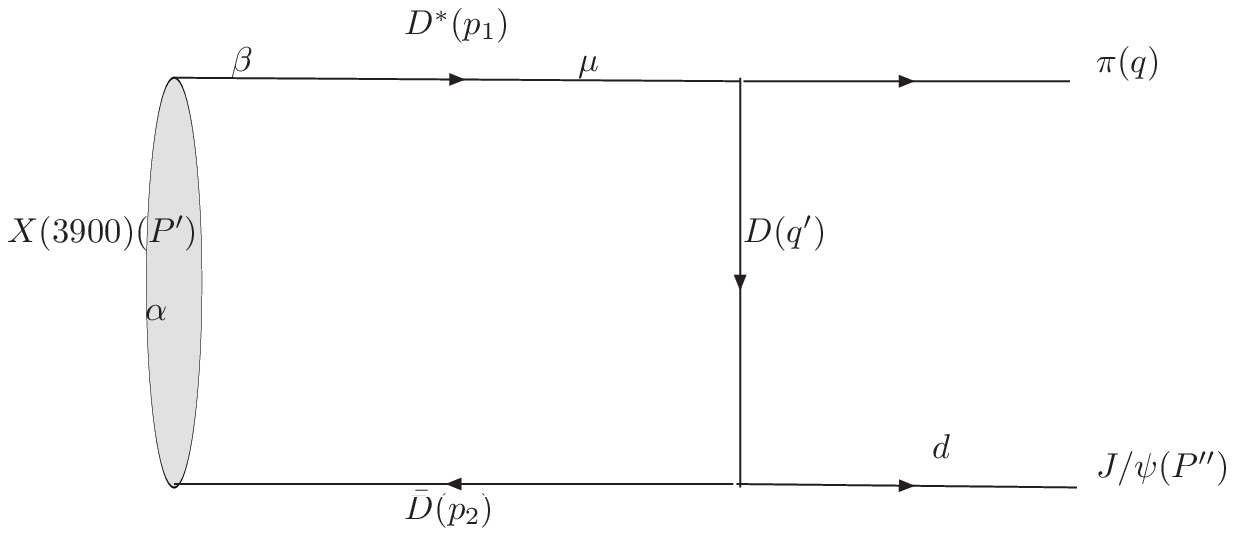}}\scalebox{0.6}{\includegraphics{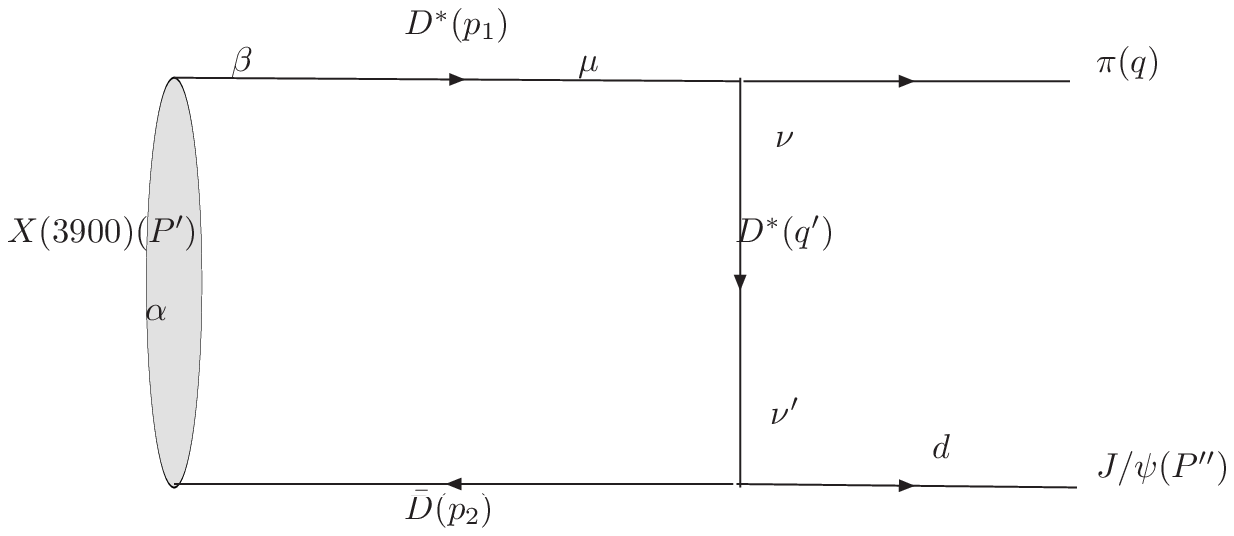}}\\
\qquad\qquad(c)\qquad\qquad\qquad\qquad\qquad\qquad\qquad
\,\,\,\,\,\,\,\,\,\,\,\,\,\,\,\,\,\,(d)\\+the Figures exchanged
the final states
\end{tabular}
\end{center}
\caption{strong decays of molecular states.}\label{fig1}
\end{figure}

The Feyman diagrams for $Z_c(3900)$ decaying into $J/\psi\pi$ by
exchanging $D$ or $D^*$  mesons are shown in Fig.\ref{fig1}. It is
noted that when calculate the rates  $Z_c(3900)\to \rho\eta_c$  or
$\psi(2s)\pi$, one can simply replace $J/\psi\pi$ by the
corresponding final states of $\rho\eta_c$  or $\psi(2s)\pi$.

Following the approach in Ref.\cite{Cheng:2003sm}, the matrix
element of diagrams  in Fig.\ref{fig1} can be cast as
\begin{eqnarray}
{\mathcal{A}}_{11}=i\frac{1}{(2\pi)^4}\int d^4
p_1\frac{[H_{A_{01}}(S^{1(a)}_{d\alpha}+S^{1(b)}_{d\alpha})+H_{A_{10}}(S^{1(c)}_{d\alpha}+S^{1(c)}_{d\alpha})]}{N_1N_1'N_2}\epsilon_1^d\epsilon^\alpha
\end{eqnarray}\label{eq1}
with
\begin{eqnarray} S_{d\alpha}^{1(b)}&&=\frac{g_{_{\psi
D^*D^*}}g_{_{\pi DD^*}}}{\sqrt{2}}g_{\alpha\beta}g^{\mu\beta}
(q_{\nu}+p_{1_{\nu}})g^{\nu\nu'}[(q'-p_2)_dg_{\mu\nu'}-q'_{\mu}g_{d\nu'}+p_{2\nu'}g_{d\mu}]
\nonumber\\&&\mathcal{F}(m_1,p_1)\mathcal{F}(m_2,p_2)\mathcal{F}^2(m_{D^*},q'),
\nonumber\\S_{d\alpha}^{1(c)}&&=\frac{-g_{_{\psi DD}}g_{_{\pi
DD^*}}}{\sqrt{2}}g_{\alpha\beta}g^{\mu\beta}
(q_\mu-q'_{\mu})(p_{2d}+q'_{d})
\mathcal{F}(m_1,p_1)\mathcal{F}(m_2,p_2)\mathcal{F}^2(m_{D},q'),
\nonumber\\S_{d\alpha}^{1(d)}&&=\frac{-g_{_{\psi DD^*}}g_{_{\pi
D^*D^*}}}{\sqrt{2}}g_{\alpha\beta}g^{\mu\beta}g^{\nu\nu'}\varepsilon_{a\mu
c\nu }p_1^a q'^{c}\varepsilon_{\omega d\nu'\rho}P''^\omega q'^\rho
\nonumber\\&&\mathcal{F}(m_1,p_1)\mathcal{F}(m_2,p_2)\mathcal{F}^2(m_{D^*},q'),
\end{eqnarray} $N_1=p_1^2-m_1^2+i\varepsilon$,
 $N_1'={q'}^2-m_{q'}^2+i\varepsilon$ and
$N_2=p_2^2-m_2^2+i\varepsilon$. However $S_{d\alpha}^{1(a)}=0$
since for strong interaction, three pseudoscalars do not couple.
The form factor
$\mathcal{F}(m_i,p^2)=\frac{(m_i+\Lambda)^2-m_i^2}{(m_i+\Lambda)^2-p^2}$
compensates the off-shell effect of the intermediate meson ($m_i$
and $p$ are the mass and momentum of the intermediate meson ). The
vertex function $H_{A}$ will be discussed later. The momentum
$p_i$  is decomposed as ($p_i^-,p_i^+,{p_i}_\perp$) in the
light-front frame. Integrating out $p_{1}^-$ with the methods
given in Ref.\cite{Cheng:1996if}  one has
\begin{eqnarray}\label{vf9.2}
\int d^4p_1 \frac{H_{A}S_{d\alpha}}{N_1N_1'N_2}{\epsilon_1}^d
{\epsilon}^{\alpha}\rightarrow-i\pi\int
dx_1d^2p_\perp\frac{h_{A}\hat S_{d\alpha}}{x_2
\hat{N_1}\hat{N_1'}} {\epsilon_1}^d {\epsilon}^{\alpha},
\end{eqnarray}
with
\begin{eqnarray*}
&&\hat{N}_1=x_1({M}^2-{M_0}^2),\\
&&\hat{N}_1^{'}=x_2q^2-x_1{M'_0}^2+x_1M''^2+2p_\perp\cdot
q_\perp,\\&&h_{A}=\sqrt{\frac{x_1x_2}{m_1'm_2}}(M'^2-M_0'^2)h_{A}'
\end{eqnarray*}
where $M$ and $M'$ represent the masses of initial and finial
mesons. The factor $\sqrt{x_1x_2}(M'^2-M_0'^2)$ in the expression
of $h_{A}$ was fixed in Ref.\cite{Cheng:2003sm} and a new factor
$\sqrt{\frac{1}{m_1m_2}}$ appears because the constituents are
bosons. $h_{A}'$ is defined in the Appendix.

To include the contributions from the zero mode ${p_1}_\mu$,
${p_1}_\nu$, ${p_1}_\mu {p_1}_\nu$ and $W_V$ in
$s_{\mu\nu}^a$
must be replaced by the appropriate expressions as discussed in
Ref.\cite{Cheng:2003sm}, for example
\begin{eqnarray}\label{eq2}
&&W_V\rightarrow w_V=M_0+m_1+m_2\nonumber\\
&&{p_1}_\mu\rightarrow\frac{x_1}{2}\mathcal{P}_\mu+(\frac{x}{2}-\frac{p_\perp\cdot
q_\perp}{q^2})q_\mu,\nonumber\\
&&\,\,\,\,\,\,\,\,\,\,\,\,\,\,......
\end{eqnarray}
with $\mathcal{P}=P'+P''$ and $P'$ and $P''$ are the momenta of the concerned mesons in the initial
and final states respectively .

More details about the derivation and some notations such as $M_0$
and  $\tilde{M}_0$ can be found in Ref.\cite{Cheng:2003sm}. With
the replacement, $h_A\hat S_{d\alpha}$ is decomposed into
\begin{eqnarray}\label{eq3}{F_{11}}
g_{d\alpha}+{F_{12}}P'_d P''_\alpha,
\end{eqnarray}
with
\begin{eqnarray}
{F_{11}}=&&\frac{g_{_{\psi D^*D^*}}g_{_{\pi DD^*}}
h_{A_{01}}}{2\sqrt{2}}(2\,{A_1^{(2)}} - 2\,{{m_1}}^2 + {{M'}}^2 +
{A_1^{(1)}}\,{{M'}}^2 + {A_2^{(1)}}\,{{M'}}^2 -
    {{M''}}^2 + 3\,{A_1^{(1)}}\,{{M''}}^2\nonumber\\&& - {A_2^{(1)}}\,{{M''}}^2 - 2\,{\hat{}N_1'} + {q^2} -
    {A_1^{(1)}}\,{q^2} - {A_2^{(1)}}\,{q^2}
)\mathcal{F}(m_1,p_1)\mathcal{F}(m_2,p_2)\mathcal{F}^2(m_{D^*},q')+\nonumber\\&&-\sqrt{2}g_{_{\psi
DD}}g_{_{\pi DD^*}}
h_{A_{10}}\,{A_1^{(2)}}\mathcal{F}(m_1,p_1)\mathcal{F}(m_2,p_2)\mathcal{F}^2(m_{D},q')+\nonumber\\&&\frac{g_{_{\psi
DD^*}}g_{_{\pi D^*D^*}} h_{A_{10}}}{2\sqrt{2}}[-2\,{A_1^{(2)}}\,(
{{M'}}^2 - {{M''}}^2 - {q^2} ) +
    ( {A_1^{(1)}} - {A_2^{(2)}} - {A_3^{(2)}} ) \,
     ( {{M'}}^4 + {( {{M''}}^2 - {q^2} ) }^2\nonumber\\&& -
       2\,{{M'}}^2\,( {{M''}}^2 + {q^2} )  )
]\mathcal{F}(m_1,p_1)\mathcal{F}(m_2,p_2)\mathcal{F}^2(m_{D^*},q')
\nonumber\\
{F_{12}}=&&\frac{g_{_{\psi D^*D^*}}g_{_{\pi
DD^*}}}{\sqrt{2}}h_{A_{01}}[1 - 6\,{A_1^{(1)}} + {A_2^{(2)}} -
2\,{A_3^{(2)}} - {A_4^{(2)}}
]\mathcal{F}(m_1,p_1)\mathcal{F}(m_2,p_2)\mathcal{F}^2(m_{D^*},q')+\nonumber\\&&-\sqrt{2}g_{_{\psi
DD}}g_{_{\pi DD^*}} h_{A_{10}}( -2 + {A_1^{(1)}} + 3\,{A_2^{(1)}}
+ {A_2^{(2)}} - 2\,{A_3^{(2)}} - {A_4^{(2)}}
)\mathcal{F}(m_1,p_1)\mathcal{F}(m_2,p_2)\nonumber\\&&\mathcal{F}^2(m_{D},q')+
\frac{g_{_{\psi DD^*}}g_{_{\pi D^*D^*}}}{\sqrt{2}}
h_{A_{10}}[2\,{A_1^{(2)}} - ( {A_1^{(1)}} - {A_2^{(2)}} -
{A_3^{(2)}} ) \,( {{M'}}^2 + {{M''}}^2 - {q^2}
)]\nonumber\\&&\mathcal{F}(m_1,p_1)\mathcal{F}(m_2,p_2)\mathcal{F}^2(m_{D^*},q'),
\end{eqnarray}\label{eq5}
where $A_{i}{^{(j)}}(i=1\sim4,j=1\sim4)$ are determined in
Ref.\cite{Cheng:2003sm}.

We define the form factors as following
\begin{eqnarray}\label{eq60}
&&{f_{11}}(m_1,m_2)=\frac{1}{32\pi^3}\int dx_2d^2p_\perp\frac{F_{11}}{x_2 \hat{N_1}\hat{N_1'}},\nonumber\\
&&{f_{12}}(m_1,m_2)=\frac{1}{32\pi^3}\int
dx_2d^2p_\perp\frac{F_{12}}{x_2 \hat{N_1}\hat{N_1'}} ,
\end{eqnarray}
which will be numerically evaluated in next section.

With these form factors   the amplitude is obtained as
\begin{eqnarray}\label{eq7}
{\mathcal{A}}_{11}&&= {f_{11}}(m_1,m_2)
\epsilon\cdot\epsilon_1+{f_{12}}(m_1,m_2) {P''\cdot\epsilon
P'\cdot\epsilon_1}.
\end{eqnarray}

The amplitude corresponding to the Feynman diagrams which are obtained by exchanging the mesons in the final states of Fig.1
can be formulated by simply exchanging $m_1$ and $m_2$.
The total amplitude is
\begin{eqnarray}\label{eq8}
\mathcal{A}_1&&=\mathcal{A}_{11}+\mathcal{A}_{12}\nonumber\\&&=[f_{11}(m_1,m_2)+f_{11}(m_2,m_1)]
\epsilon\cdot\epsilon_1+[f_{12}(m_1,m_2)+f_{12}(m_2,m_1)]
{P''\cdot\epsilon P'\cdot\epsilon_1}\nonumber\\&&=g_{11}
\epsilon\cdot\epsilon_1+g_{12} {P''\cdot\epsilon
P'\cdot\epsilon_1}.
\end{eqnarray}

\subsection{$Z_c(3900)\rightarrow  D\bar D^*$}
\begin{figure}
\begin{center}
\begin{tabular}{ccc}
\scalebox{0.6}{\includegraphics{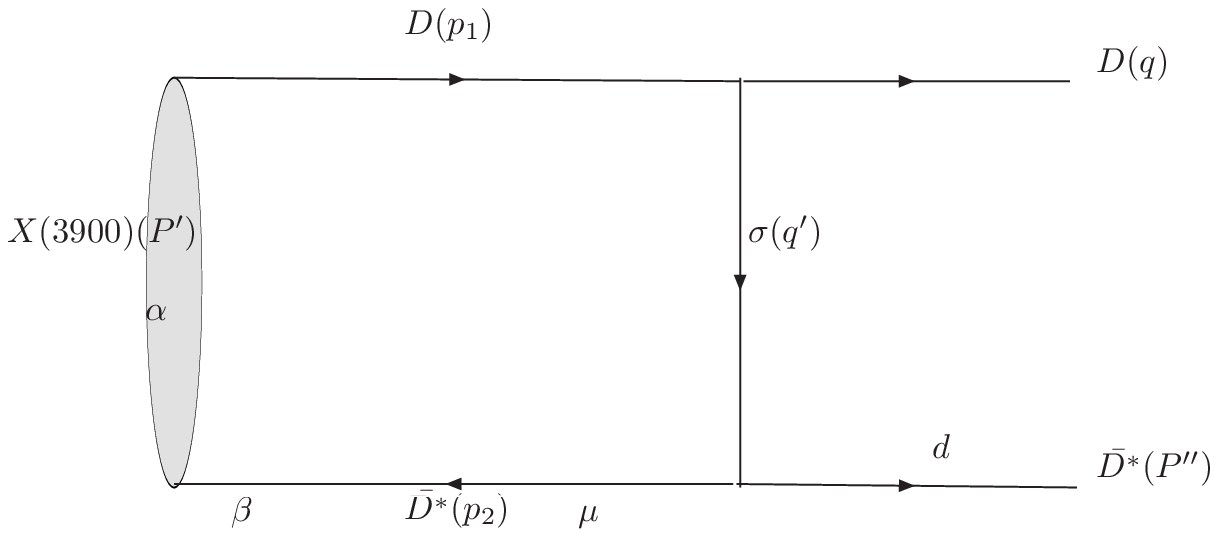}}\scalebox{0.6}{\includegraphics{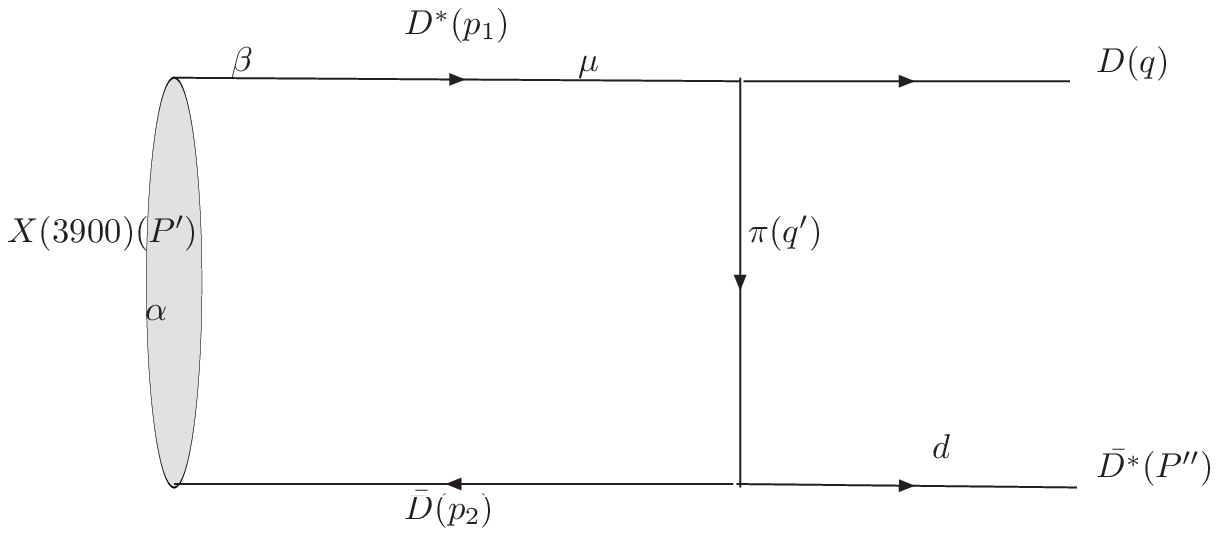}}
\\
\qquad\qquad(I)\qquad\qquad\qquad\qquad\qquad\qquad\qquad
\,\,\,\,\,\,\,\,\,\,\,\,\,\,\,\,\,\,(II)\end{tabular}
\end{center}
\caption{strong decay of molecular state.}\label{Fig2}
\end{figure}

The corresponding Feynman diagrams are shown in Fig.\ref{Fig2}.
Generally the intermediate mesons should include $\rho$, $\omega$,
$\pi$ and $\sigma$. However for an $I=1$ molecular state the
contributions from $\rho$ and $\omega$  nearly
cancel each other as discussed in Ref.\cite{Guo:2007mm}. In terms
of the vertex function presented in the attached appendix, the hadronic matrix element corresponding to
the  diagrams in Fig.\ref{Fig2} is written as
\begin{eqnarray}\label{eq9}
\mathcal{A}_2=i\frac{1}{(2\pi)^4}\int d^4 p_1\frac{H_{A_{01}}
S^{2(a)}_{d\alpha}+H_{A_{10}}S^{2(b)}_{d\alpha}}{N_1N_1'N_2}\epsilon_1^d\epsilon^\alpha
\end{eqnarray}
with \begin{eqnarray}&&S^{2(a)}_{d\alpha}=\frac{g_{_{\sigma
DD}}g_{_{\sigma
D^*D^*}}}{\sqrt{2}}g_{\alpha\beta}g^{\mu\beta}g_{d\mu}
\mathcal{F}(m_1,p_1)\mathcal{F}(m_2,p_2)\mathcal{F}^2(m_{\sigma},q'),
\nonumber\\&&S_{d\alpha}^{2(b)}=\frac{-g_{_{\pi DD^*}}g_{_{\pi
DD^*}}}{\sqrt{2}}g_{\alpha\beta}g^{\mu\beta}
(-q_\mu+q'{\mu})(p_{2d}+q'_{d})
\mathcal{F}(m_1,p_1)\mathcal{F}(m_2,p_2)\mathcal{F}^2(m_{D},q').\end{eqnarray}
Carrying out the integral, $h_A\hat S_{d\alpha}$ is decomposed
into
\begin{eqnarray}\label{eq10}{F_{21}}
g_{d\alpha}+{F_{22}}P'_d P''_\alpha,
\end{eqnarray}
with
\begin{eqnarray}\label{eq11}
F_{21}&&=\frac{g_{_{\sigma DD}}g_{_{\sigma
D^*D^*}}}{\sqrt{2}}\mathcal{F}(m_1,p_1)\mathcal{F}(m_2,p_2)\mathcal{F}^2(m_{\sigma},q')+\sqrt{2}g_{_{\pi
DD^*}}g_{_{\pi
DD^*}}\,{A_1^{(2)}}\nonumber\\&&\mathcal{F}(m_1,p_1)\mathcal{F}(m_2,p_2)\mathcal{F}^2(m_{\pi},q')\nonumber\\
F_{22}&&=\sqrt{2}g_{_{\pi DD^*}}g_{_{\pi DD^*}}\,( -2 +
{A_1^{(1)}} + 3\,{A_2^{(1)}} + {A_2^{(2)}} - 2\,{A_3^{(2)}} -
{A_4^{(2)}}
)\nonumber\\&&\mathcal{F}(m_1,p_1)\mathcal{F}(m_2,p_2)\mathcal{F}^2(m_{\pi},q').
\end{eqnarray}

Similar to the definitions in Eq.\ref{eq60} the amplitude then is
\begin{eqnarray}\label{eq12}
\mathcal{A}_2&&=f_{21} \epsilon\cdot\epsilon_1+f_{22}
{P''\cdot\epsilon P'\cdot\epsilon_1}.
%\nonumber\\&&=g_{21}
%\epsilon\cdot\epsilon_1+g_{22} {P''\cdot\epsilon
%P'\cdot\epsilon_1}.
\end{eqnarray}

\section{The strong decays of $Z_c(4030)$ which is assumed to be a $1^{+}$ molecular state of $D^*\bar D^*$ }
Now let us consider strong decays of  $Z_c(4030)$ which is  assumed to be  a $D^*\bar D^*$ molecular state.
Similar to what we have done for $Z_c(3900)$, we calculate the decay rates for $Z_c(4030)\rightarrow J/\psi\pi$,
$Z_c(4030)\rightarrow \psi'\pi$, $Z_c(4030)\rightarrow
\rho\eta_c$,  $Z_c(4030)\rightarrow D\bar D^*$, and one more channel:
$Z_c(4030)\rightarrow D^*\bar D^*$ which is open at the energy of 4030 MeV.

\begin{figure}
\begin{center}
\begin{tabular}{ccc}
\scalebox{0.6}{\includegraphics{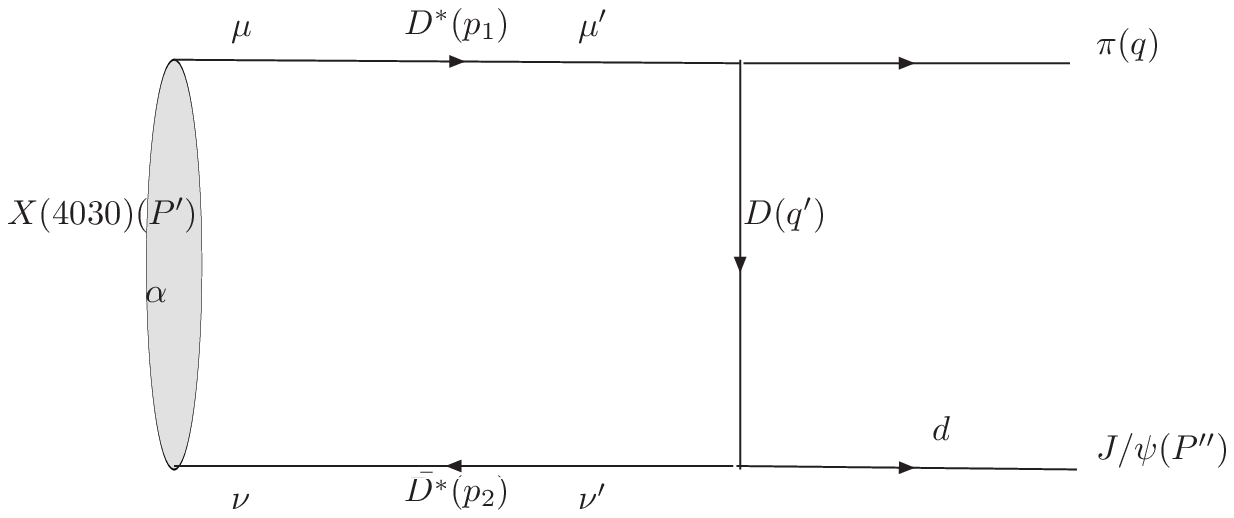}}\scalebox{0.6}{\includegraphics{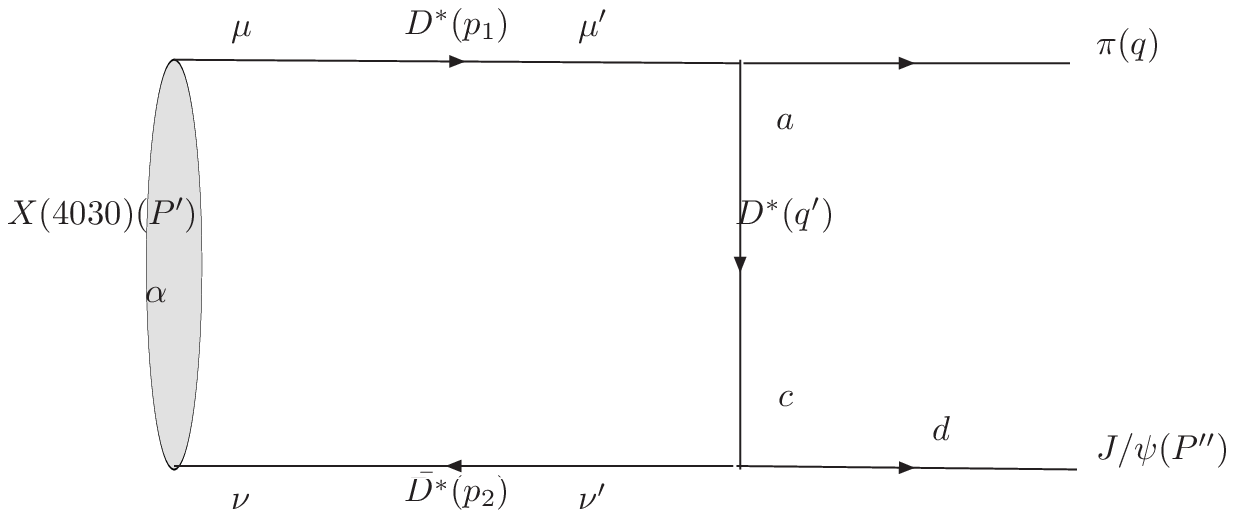}}\\
\qquad\qquad(a)\qquad\qquad\qquad\qquad\qquad\qquad\qquad
\,\,\,\,\,\,\,\,\,\,\,\,\,\,\,\,\,\,(b)\\+the Figures exchanged
the final states
\end{tabular}
\end{center}
\caption{strong decay  $Z_c(4030)\rightarrow J/\psi\pi
$}\label{Fig3}
\end{figure}
\subsection{$Z_c(4030)\rightarrow J/\psi\pi\,(\rho\eta_c \,{\rm or} \,\psi'\pi)$}
The Feynman diagrams are shown in Fig.\ref{Fig3}. In terms of the vertex function given in the appendix,
the hadronic matrix element is
\begin{eqnarray}
\mathcal{A}_{31}=i\frac{1}{(2\pi)^4}\int d^4
p_1\frac{H_{A_1}}{N_1N_1'N_2}(S^{3(a)}_{d\alpha}+S^{3(b)}_{d\alpha})\epsilon_1^d\epsilon^\alpha,
\end{eqnarray}\label{eq13}
where
$$S_{d\alpha}^{3(a)}=g_{_{\psi DD^*}}g_{_{\pi DD^*}}\varepsilon_{\mu\nu\alpha\beta}
g^{\mu\mu'}(2q_{\mu'}-p_{1\mu'})P'^{\beta}g^{\nu\nu'}\varepsilon_{ab\nu'd}P''^a
p_2^{b}
\mathcal{F}(m_1,p_1)\mathcal{F}(m_2,p_2)\mathcal{F}^2(m_{D},q'),$$
and $S_{d\alpha}^{3(b)}=g_{_{\psi D^*D^*}}g_{_{\pi
D^*D^*}}\varepsilon_{\mu\nu\alpha\beta}g^{\mu\mu'}P'^{\beta}g^{\nu\nu'}\varepsilon_{\omega\mu'\rho
a}
p_1^{\omega}q'^{\rho}g^{ac}[(q'-p_2)_dg_{c\nu'}-q'_{\nu}g_{dc}+p_{2c}g_{d\nu'}]$
$\mathcal{F}(m_1,p_1)\mathcal{F}(m_2,p_2)\mathcal{F}^2(m_{D^*},q').$
Carrying out the integration and making the required replacements,
we have
\begin{eqnarray}\label{eq14}
h_{A_1}(\hat S^{3(a)}_{d\alpha}+\hat S^{3(b)}_{d\alpha})={F_{31}}
g_{d\alpha}+{F_{32}}P'_d P''_\alpha,
\end{eqnarray}
with
\begin{eqnarray}\label{eq15}
{F_{31}}=&&\frac{g_{_{\psi DD^*}}g_{_{\pi DD^*}}h_{A_1}}{4}
      [ 2\,{{M'}}^4 - {A_1^{(1)}}\,{{M'}}^4 - 3\,{A_2^{(1)}}\,{{M'}}^4 +
    3\,{A_2^{(2)}}\,{{M'}}^4 + 4\,{A_3^{(2)}}\,{{M'}}^4 + {A_4^{(2)}}\,{{M'}}^4\nonumber\\&& -
    4\,{{M'}}^2\,{{M''}}^2 + 2\,{A_1^{(1)}}\,{{M'}}^2\,{{M''}}^2 +
    6\,{A_2^{(1)}}\,{{M'}}^2\,{{M''}}^2 + 10\,{A_2^{(2)}}\,{{M'}}^2\,{{M''}}^2 -
    4\,{A_3^{(2)}}\,{{M'}}^2\,{{M''}}^2 \nonumber\\&&- 2\,{A_4^{(2)}}\,{{M'}}^2\,{{M''}}^2 +
    2\,{{M''}}^4 - {A_1^{(1)}}\,{{M''}}^4 - 3\,{A_2^{(1)}}\,{{M''}}^4 +
    3\,{A_2^{(2)}}\,{{M''}}^4 + {A_4^{(2)}}\,{{M''}}^4 \nonumber\\&&+
    4\,{A_1^{(2)}}\,( {{M'}}^2 + {{M''}}^2 - {q^2} ) -
    2\,{{m_1}}^2\,( {{M'}}^2 + {{M''}}^2 - {q^2} )  -
    4\,{{M'}}^2\,{q^2} + 2\,{A_1^{(1)}}\,{{M'}}^2\,{q^2} \nonumber\\&&+
    6\,{A_2^{(1)}}\,{{M'}}^2\,{q^2} - 4\,{A_2^{(2)}}\,{{M'}}^2\,{q^2} -
    4\,{A_3^{(2)}}\,{{M'}}^2\,{q^2} - 4\,{{M''}}^2\,{q^2} +
    2\,{A_1^{(1)}}\,{{M''}}^2\,{q^2} \nonumber\\&&+ 6\,{A_2^{(1)}}\,{{M''}}^2\,{q^2} -
    4\,{A_2^{(2)}}\,{{M''}}^2\,{q^2} + 2\,{{q^4}} - {A_1^{(1)}}\,{{q^4}} -
    3\,{A_2^{(1)}}\,{{q^4}} + {A_2^{(2)}}\,{{q^4}} - {A_4^{(2)}}\,{{q^4}} \nonumber\\&&-
    2\,{{M'}}^2\,( -{{M_0'}}^2 + {{M'}}^2 ) \,{x_1} +
    2\,( {{M_0'}}^2 - {{M'}}^2 ) \,{{M''}}^2\,{x_1} +
    2\,( -{{M_0'}}^2 + {{M'}}^2 ) \,{q^2}\,{x_1} ]\nonumber\\&&
    \mathcal{F}(m_1,p_1)\mathcal{F}(m_2,p_2)\mathcal{F}^2(m_{D},q')
\nonumber\\&&+\frac{g_{_{\psi D^*D^*}}g_{_{\pi
D^*D^*}}h_{A_1}}{2}\{6\,{A_1^{(2)}}\,( {{M'}}^2 - {{M''}}^2 +
{q^2} ) -
    ( {A_1^{(1)}} + {A_2^{(2)}} ) \,
     [ {{M'}}^4 + {( {{M''}}^2 - {q^2} ) }^2\nonumber\\&& -
       2\,{{M'}}^2\,( {{M''}}^2 + {q^2} )  ]\}\mathcal{F}(m_1,p_1)\mathcal{F}(m_2,p_2)\mathcal{F}^2(m_{D^*},q')
\nonumber\\
{F_{32}}=&&\frac{g_{_{\psi DD^*}}g_{_{\pi
DD^*}}h_{A_1}}{2}[-4\,{A_1^{(2)}} + 2\,{{m_1}}^2 - 2\,{{M'}}^2 +
{A_1^{(1)}}\,{{M'}}^2 +
    3\,{A_2^{(1)}}\,{{M'}}^2 - 3\,{A_2^{(2)}}\,{{M'}}^2 \nonumber\\&&- 4\,{A_3^{(2)}}\,{{M'}}^2 -
    {A_4^{(2)}}\,{{M'}}^2 - 2\,{{M''}}^2 + {A_1^{(1)}}\,{{M''}}^2 +
    3\,{A_2^{(1)}}\,{{M''}}^2 - 3\,{A_2^{(2)}}\,{{M''}}^2 \nonumber\\&&- {A_4^{(2)}}\,{{M''}}^2 +
    2\,{q^2} - {A_1^{(1)}}\,{q^2} - 3\,{A_2^{(1)}}\,{q^2} + {A_2^{(2)}}\,{q^2} -
    {A_4^{(2)}}\,{q^2} - 2\,{{M_0'}}^2\,{x_1} + 2\,{{M'}}^2\,{x_1}
]\nonumber\\&&\mathcal{F}(m_1,p_1)\mathcal{F}(m_2,p_2)\mathcal{F}^2(m_{D},q')\nonumber\\
   && +g_{_{\psi D^*D^*}}g_{_{\pi D^*D^*}}h_{A_1}[6\,{A_1^{(2)}} + {A_1^{(1)}}\,( -7\,{{M'}}^2 + {{M''}}^2 - {q^2} )  +
  {A_2^{(2)}}\,( 9\,{{M'}}^2 + {{M''}}^2 - {q^2} )]
  \nonumber
  \\&&\mathcal{F}(m_1,p_1)\mathcal{F}(m_2,p_2)\mathcal{F}^2(m_{D^*},q').
\end{eqnarray}

We define the form factors as following
\begin{eqnarray}\label{eq16}
&&{f_{31}}(m_1,m_2)=\frac{1}{32\pi^3}\int dx_2d^2p_\perp\frac{F_{31}}{x_2 \hat{N_1}\hat{N_1'}},\nonumber\\
&&{f_{32}}(m_1,m_2)=\frac{1}{32\pi^3}\int
dx_2d^2p_\perp\frac{F_{32}}{x_2 \hat{N_1}\hat{N_1'}} ,
\end{eqnarray}
which will be numerically evaluated in next section.

With these form factors   the amplitude is obtained as
\begin{eqnarray}\label{eq17}
\mathcal{A}_{31}&&= {f_{31}}(m_1,m_2)
\epsilon\cdot\epsilon_1+{f_{32}}(m_1,m_2) {P''\cdot\epsilon
P'\cdot\epsilon_1}.
\end{eqnarray}

Similarly, the amplitude corresponding the Feynman diagrams which are obtained by switching around the mesons in the final states
can be easily obtained by exchanging $m_1$ and $m_2$.
The total
amplitude is
\begin{eqnarray}\label{eq18}
\mathcal{A}_3&&=\mathcal{A}_{31}+\mathcal{A}_{32}\nonumber\\&&=[f_{31}(m_1,m_2)+f_{31}(m_2,m_1)]
\epsilon\cdot\epsilon_1+[f_{32}(m_1,m_2)+f_{32}(m_2,m_1)]
{P''\cdot\epsilon P'\cdot\epsilon_1}\nonumber\\&&=g_{31}
\epsilon\cdot\epsilon_1+g_{32} {P''\cdot\epsilon
P'\cdot\epsilon_1}.
\end{eqnarray}

\subsection{$Z_c(4030)\rightarrow  D\bar D^*$}
Since there does not exist an effective coupling of
$\sigma$ with a psedoscalar and a vector, only the Feynman diagram
with intermediate $\pi$ can contribute.

\begin{figure}
\begin{center}
\begin{tabular}{ccc}
\scalebox{0.6}{\includegraphics{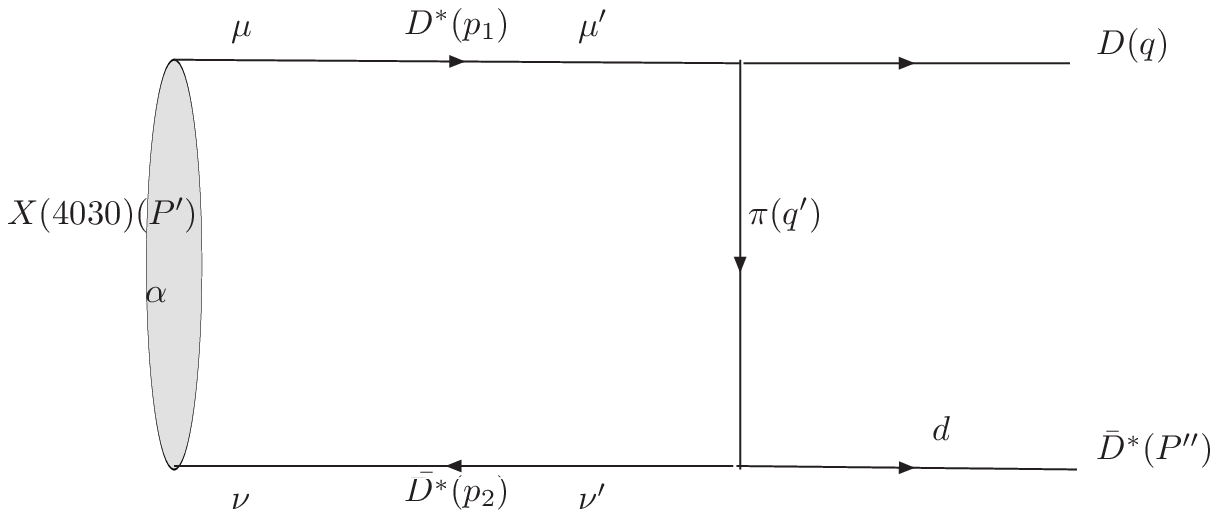}}
\end{tabular}
\end{center}
\caption{strong decay of molecular state.}\label{fig4}
\end{figure}

In terms of the vertex function, the hadronic matrix element corresponding to (a) and (b) of Fig.\ref{fig4} is
\begin{eqnarray}\label{eq19}
\mathcal{A}_4=i\frac{1}{(2\pi)^4}\int d^4
p_1\frac{H_{A_1}}{N_1N_1'N_2}S_{d\alpha}\epsilon_1^d\epsilon^\alpha
\end{eqnarray}
where
$$S_{d\alpha}=-g_{_{\pi DD^*}}g_{_{\pi D^*D^*}}\varepsilon_{\mu\nu\alpha\beta}g^{\mu\mu'}
(-2q_{\mu'}+p_{1\mu'})P'^{\beta}g^{\nu\nu'}\varepsilon_{ab\nu'd}P''^ap_2^{b}
\mathcal{F}(m_1,p_1)\mathcal{F}(m_2,p_2)\mathcal{F}^2(m_{\pi},q'),$$
Carrying out the integration and making the replacements, we have
\begin{eqnarray}\label{eq20}
\hat S_{d\alpha}={F_{41}} g_{d\alpha}+{F_{42}}P'_d P''_\alpha,
\end{eqnarray}
with \begin{eqnarray} {F_{41}}=&& \frac{g_{_{\pi DD^*}}g_{_{\pi
D^*D^*}}h_{A_1}}{4}[-2\,{{M'}}^4 + {A_1^{(1)}}\,{{M'}}^4 +
3\,{A_2^{(1)}}\,{{M'}}^4 -
    3\,{A_2^{(2)}}\,{{M'}}^4 - 4\,{A_3^{(2)}}\,{{M'}}^4 - {A_4^{(2)}}\,{{M'}}^4 \nonumber\\&&+
    4\,{{M'}}^2\,{{M''}}^2 - 2\,{A_1^{(1)}}\,{{M'}}^2\,{{M''}}^2 -
    6\,{A_2^{(1)}}\,{{M'}}^2\,{{M''}}^2 -
    10\,{A_2^{(2)}}\,{{M'}}^2\,{{M''}}^2 +
    4\,{A_3^{(2)}}\,{{M'}}^2\,{{M''}}^2\nonumber\\&& +
    2\,{A_4^{(2)}}\,{{M'}}^2\,{{M''}}^2 - 2\,{{M''}}^4 +
    {A_1^{(1)}}\,{{M''}}^4 + 3\,{A_2^{(1)}}\,{{M''}}^4 -
    3\,{A_2^{(2)}}\,{{M''}}^4 - {A_4^{(2)}}\,{{M''}}^4\nonumber\\&&  -
    4\,{A_1^{(2)}}\,( {{M'}}^2 + {{M''}}^2 - {q^2} ) +
    2\,{{m_1'}}^2\,( {{M'}}^2 + {{M''}}^2 - {q^2} )  +
    4\,{{M'}}^2\,{q^2} - 2\,{A_1^{(1)}}\,{{M'}}^2\,{q^2}\nonumber\\&& -
    6\,{A_2^{(1)}}\,{{M'}}^2\,{q^2} + 4\,{A_2^{(2)}}\,{{M'}}^2\,{q^2} +
    4\,{A_3^{(2)}}\,{{M'}}^2\,{q^2} + 4\,{{M''}}^2\,{q^2} -
    2\,{A_1^{(1)}}\,{{M''}}^2\,{q^2} \nonumber\\&&- 6\,{A_2^{(1)}}\,{{M''}}^2\,{q^2} +
    4\,{A_2^{(2)}}\,{{M''}}^2\,{q^2} - 2\,q^4 + {A_1^{(1)}}\,q^4 +
    3\,{A_2^{(1)}}\,q^4\nonumber\\&& - {A_2^{(2)}}\,q^4 + {A_4^{(2)}}\,q^4 +
    2\,( -{{M_0'}}^2 + {{M'}}^2 ) \,
     ( {{M'}}^2 + {{M''}}^2 - {q^2} ) \,{x_1}]\mathcal{F}(m_1,p_1)\mathcal{F}(m_2,p_2)\mathcal{F}^2(m_{\pi},q')
\nonumber\\{F_{42}}=&&\frac{g_{_{\pi DD^*}}g_{_{\pi
D^*D^*}}h_{A_1}}{2}[4\,{A_1^{(2)}} - 2\,{{m_1}}^2 +
    ( 2 - {A_1^{(1)}} - 3\,{A_2^{(1)}} + 3\,{A_2^{(2)}} + {A_4^{(2)}} ) \,{{M''}}^2 +
    ( -2 + {A_1^{(1)}} + \nonumber\\&& 3\,{A_2^{(1)}} - {A_2^{(2)}} + {A_4^{(2)}} ) \,{q^2} +
    {{M'}}^2\,( 2 - {A_1^{(1)}} - 3\,{A_2^{(1)}} + 3\,{A_2^{(2)}}  + 4\,{A_3^{(2)}} +
       {A_4^{(2)}} - 2\,{x_1} ) \nonumber\\&& + 2\,{{M_0'}}^2\,{x_1}]\mathcal{F}(m_1,p_1)\mathcal{F}(m_2,p_2)\mathcal{F}^2(m_{\pi},q').
\end{eqnarray}\label{eq21}

The amplitude can be eventually reached as
\begin{eqnarray}\label{eq22}
\mathcal{A}_4&&=f_{41} \epsilon\cdot\epsilon_1+f_{42}
{P''\cdot\epsilon P'\cdot\epsilon_1}.
%\nonumber\\&&=g_{41}
%\epsilon\cdot\epsilon_1+g_{42} {P''\cdot\epsilon
%P'\cdot\epsilon_1}.
\end{eqnarray}

\subsection{$Z_c(4030)\rightarrow  D^*\bar D^*$}
For this decay mode there are two Feynman diagrams which are
induced by exchanging $\pi$ and $\sigma$ between the two
constituents $D^*$ and $\bar D^*$, making contribution.

\begin{figure}
\begin{center}
\begin{tabular}{ccc}
\scalebox{0.6}{\includegraphics{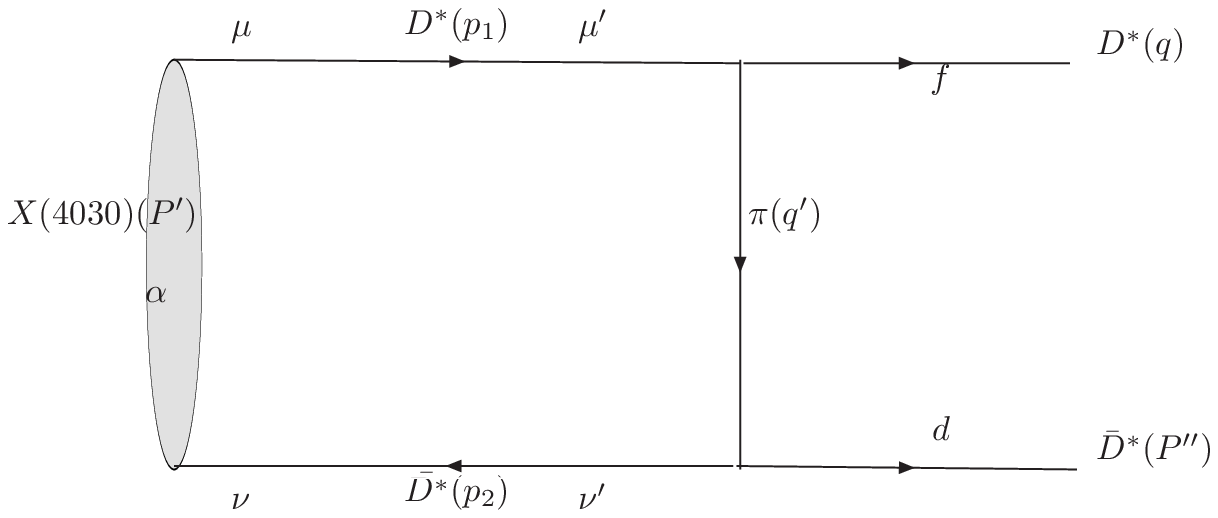}}\scalebox{0.6}{\includegraphics{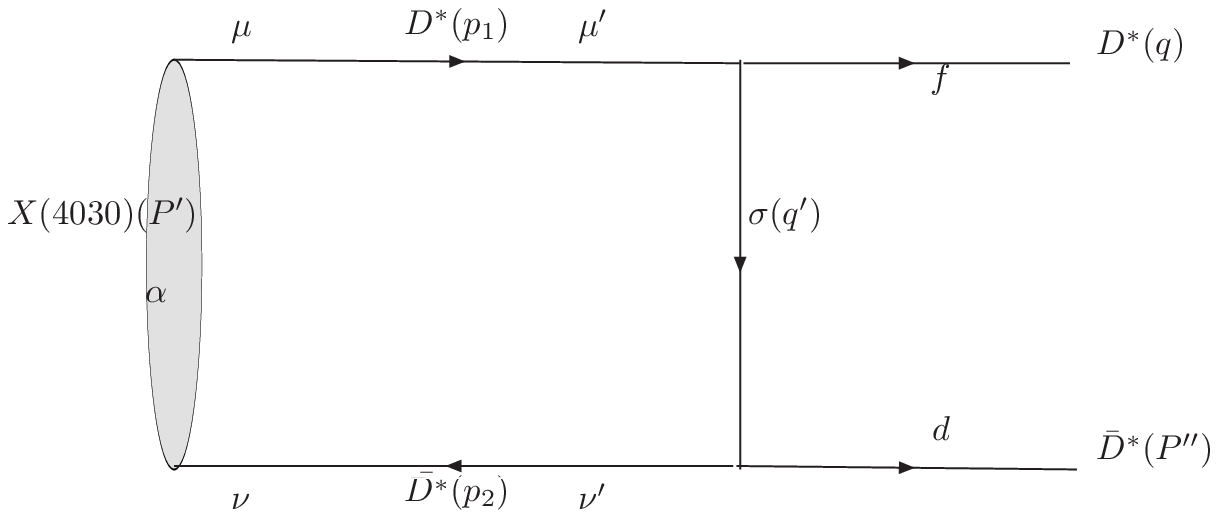}}
\\
\qquad\qquad(a)\qquad\qquad\qquad\qquad\qquad\qquad\qquad
\,\,\,\,\,\,\,\,\,\,\,\,\,\,\,\,\,\,(b)\end{tabular}
\end{center}
\caption{strong decay of molecular state.}\label{Fig5}
\end{figure}

In terms of the vertex function, the hadronic matrix element
corresponding to diagrams (a) and (b) of Fig.\ref{Fig5} is
\begin{eqnarray}\label{eq24}
\mathcal{A}_5=i\frac{1}{(2\pi)^4}\int d^4
p_1\frac{H_{A_1}}{N_1N_1'N_2}(S^{5(a)}_{df\alpha}+S^{5(b)}_{df\alpha})\epsilon_1^d\epsilon_2^f\epsilon^\alpha
\end{eqnarray}
\begin{eqnarray}
&&S^{5(a)}_{df\alpha}=g_{_{\pi D^*D^*}}g_{_{\pi
D^*D^*}}\varepsilon_{\mu\nu\alpha\beta}\varepsilon_{\rho\mu'\omega
f} \varepsilon_{a d b\nu'}g^{\mu\mu'}g^{\nu\nu'}P'^{\beta}
(p_{1\rho})q_\nu P''^ap_2^{b}
\mathcal{F}(m_1,p_1)\mathcal{F}(m_2,p_2)\mathcal{F}^2(m_{\pi},q'),
\nonumber\\&&S^{5(b)}_{df\alpha}=g_{_{\sigma D^*D^*}}g_{_{\sigma
D^*D^*}}\varepsilon_{\mu\nu\alpha\beta}
g^{\mu\mu'}g^{\nu\nu'}P'^{\beta} g^{f\mu'}g^{d\nu'}
\mathcal{F}(m_1,p_1)\mathcal{F}(m_2,p_2)\mathcal{F}^2(m_{\pi},q').
\end{eqnarray}
carrying the integration and making the replacement, one has
\begin{eqnarray}\label{eq25}
\hat S^{5(a)}_{df\alpha}+\hat S^{5(b)}_{df\alpha}={F_{51}}
\varepsilon_{df\mu\nu}P'^\mu P''^\nu P''^\alpha+{F_{52}}
\varepsilon_{df\alpha\beta}P'_\beta,
\end{eqnarray}
with \begin{eqnarray} {F_{51}}=&& 2g_{_{\pi D^*D^*}}g_{_{\pi
D^*D^*}}h_{A_1} A_1^1 (A_1^1 + A_2^1 - 1)
\mathcal{F}(m_1,p_1)\mathcal{F}(m_2,p_2)\mathcal{F}^2(m_{\pi},q')
\nonumber\\{F_{52}}=&&g_{_{\sigma D^*D^*}}g_{_{\sigma
D^*D^*}}h_{A_1}\mathcal{F}(m_1,p_1)\mathcal{F}(m_2,p_2)\mathcal{F}^2(m_{\sigma},q').
\end{eqnarray}\label{eq26}

The amplitude is
\begin{eqnarray}\label{eq27}
\mathcal{A}_5&&={f_{51}} \varepsilon_{df\mu\nu}P'^\mu P''^\nu
P''^\alpha \epsilon_\alpha\epsilon_{1d}\epsilon_{2f}+{f_{52}}
\varepsilon_{df\alpha\beta}P'_\beta\epsilon_\alpha\epsilon_{1d}\epsilon_{2f}.
%\nonumber\\&&=g_{41}
%\epsilon\cdot\epsilon_1+g_{42} {P''\cdot\epsilon
%P'\cdot\epsilon_1},
\end{eqnarray}
where the expressions of $f_{51}$ and $f_{52}$ are similar to
Eq.\ref{eq60} and  Eq.\ref{eq16}.

\section{numerical results}
In this section we present our theoretical predictions on the
decay rates of the concerned modes. The key point is to calculate
the corresponding form factors we deduced in last section. Those
formulas involve  many parameters which need to be priori fixed.
We use the BES dada 3.899 GeV\cite{Ablikim:2013mio} as the mass of
$Z_c(3900)$ and the mass of  $Z_c(4030)$ is assigned to be 4.03
GeV. The masses of the decay products and intermediate mesons are
set as $m_{J/\psi}=3.096$ GeV, $m_{\psi'}=3.686$ GeV,
$m_{\eta_c}=2.981$ GeV, $m_{\rho}=0.77$ GeV, $m_{\pi}=0.139$ GeV,
$m_{D}=1.869$ GeV, $m_{D^*}=2.007$ GeV and $m_\sigma=0.5$ GeV
taken from Ref.\cite{PDG12}. In
Refs.\cite{Haglin:1999xs,Oh:2000qr} the coupling constants
$g_{_{\pi DD^*}}$ and $g_{_{\pi D^*D^*}}$ were fixed to be 8.8 and
9.08 GeV$^{-1}$ respectively.  For the coupling of $\psi
D^{(*)}D^{(*)}$, $\psi DD^{*}$ and $\psi D^{*}D^{*}$ there exists
a simple, but approximate relation $g_{_{\psi DD}}=m_Dg_{_{\psi
DD^*}}=g_{_{\psi D^*D^*}}$\cite{Deandrea:2003pv,Meng:2007cx} and
$g_{_{\psi DD}}=g_{_{\psi D^*D^*}}=8.0$\cite{Lin:1999ad}, so we
can fix $g_{_{\psi DD^*}}=4.28$ GeV$^{-1}$. In the heavy quark
limit the relation $g_{_{\eta_c D^*D}}=m_Dg_{_{\eta_c
D^*D^*}}=g_{_{\psi DD}}$ should exist. The coupling constant
$g_{_{\rho DD}}$ and $g_{_{\rho D^*D^*}}$ are set to be
3\cite{Guo:2007mm} in our calculation and $g_{_{\rho DD^*}}=3$
GeV$^{-1}$ is adopted. $g_{_{\sigma D^*D^*}}=2m_{D^*}g_\sigma$ and
$g_{_{\sigma DD}}=2m_{D}g_\sigma$ with
$g_\sigma=0.76$\cite{Lee:2009hy} are also reasonable
approximations. $\Lambda$ in the vertex $\mathcal{F}$ is a cutoff
parameter which was suggested to set as 0.88 GeV to 1.1 GeV in
Ref.\cite{Meng:2007cx} and we will use both of the values in our
calculation and compare the results. The other parameter $\beta$
in the wavefunction is not very clear so far and its value is
estimated to be near or smaller than the number of $\beta$ for the
wavefunction of $J/\psi$ which is fixed to be 0.631 GeV$^{-1}$ in
Ref.\cite{Ke:2011jf}.
% To see how it affects the final
%theoretical predictions on the decay rates for all the concerned
%modes, we deliberately vary it and plot the dependence of the decay rates of
%$Z_c(3900)\to J/\psi\pi$ and $Z_c(4030)\to J/\psi\pi$ on $\beta$.

Since we  derive the  form factors in the frame of $q^+=0$ (
$q^2<0$) i.e. in space-like region, we extend these form factors
to the time-like region according to the normal procedure provided
in literatures. Then letting $q^2$ take the value of $m^2$ , the
physical form factors are obtained. In Ref.\cite{Cheng:2003sm} a
three-parameter form was suggested as
\begin{eqnarray}\label{eq23}
 F(q^2)=\frac{F(0)}{
  \left[1-a\left(\frac{q^2}{M_{\Lambda_b}^2}\right)
  +b\left(\frac{q^2}{M_{\Lambda_b}^2}\right)^2\right]}.
 \end{eqnarray}

However we find the form Eq.\ref{eq23} does not fit the numerical
values satisfactorily (see the figures), so instead, we employ a
polynomial
\begin{eqnarray}\label{eq24}
 F(q^2)=F(0)
  \left[1+a\left(\frac{q^2}{M_{\Lambda_b}^2}\right)
  +b\left(\frac{q^2}{M_{\Lambda_b}^2}\right)^2+c\left(\frac{q^2}{M_{\Lambda_b}^2}\right)^3+d\left(\frac{q^2}{M_{\Lambda_b}^2}\right)^4\right].
 \end{eqnarray}

The resultant form factors and the effective coupling constants
are listed in table \ref{Tab:t1}.

\begin{table}[!h]
\caption{The  form factors given in the
five-parameter form ($\Lambda=0.88$ GeV, $\beta=0.631$ GeV$^{-1}$). The numbers above the horizontal line
correspond to $Z_c(3900)$ and those numbers below the line correspond to $Z_c(4030)$. The definitions of
the form factors are given in last section.}\label{Tab:t1}
\begin{ruledtabular}
\begin{tabular}{cccccc|cccccc}
  $F$    & $F(0)$
& $a$  &  $b$ &$c$&$d$& $F$& $F(0)$ &  $a$  & $b$&$c$&$d$\\\hline
  $g_{11}^{J/\psi\pi}$  &   -1.05      &   10.59    &  17.83&14.73&4.90& $g_{12}^{J/\psi\pi}$ &-1.16 &  2.33  &  3.21&2.51&0.82  \\
  $g_{11}^{\psi(2S)\pi}$  &   -4.24      &   9.32    &  18.271&16.57&5.82& $g_{12}^{J/\psi(2S)\pi}$ &-4.35 &  3.18  & 5.33&4.67&1.63 \\
  $g_{11}^{\rho\eta_c}$  &   0.065      & - 6.81    &  -9.34&-6.769&-2.10& $g_{12}^{J/\rho\eta_c}$ &-0.083 &  1.66  &  1.85&1.30&0.40 \\
   $f_{21}$  & 0.052      & 5.34     &12.57 &13.43&5.27& $f_{22}$& 0.39 &  5.72  & 13.89&15.12&5.99   \\\hline
$g_{31}^{J/\psi\pi}$  &  2.90     &   -10.39    &  -21.90&-20.54&-7.53& $g_{32}^{J/\psi\pi}$ &0.64 &  3.23  & 5.05&4.26&1.49  \\
  $g_{31}^{\psi(2S)\pi}$  &   -0.84      &   -59.48    &-148.71&-155.46&-60.64& $g_{32}^{\psi(2S)\pi}$ &-1.30 &  3.89  &  6.86&6.20&2.25 \\
  $g_{31}^{\rho\eta_c}$  &   -0.46     & 0.15    & -0.16&-0.17&-0.06& $g_{32}^{J/\rho\eta_c}$ &-0.011 & 5.80  &  8.78&6.99&2.34  \\
    $f_{41}$  &      -0.28   &    0.031  & -4.79&-7.81&-3.76& $f_{42}$& -0.26 &  5.25  &
  12.676&14.20&5.89\\   $f_{51}$  &      0.074   &    5.35  & 12.91&14.42&5.96& $f_{52}$& -$0.053$ &  5.14  &
 12.04&13.23&5.42
\end{tabular}
\end{ruledtabular}
\end{table}

\begin{table}[!h]
\caption{The decay widths of some modes ($\Lambda=0.88$ GeV,
$\beta=0.631$ GeV$^{-1}$).}\label{Tab:t2}
\begin{ruledtabular}
\begin{tabular}{cc|cc}
  decay mode   &  width(MeV)&  decay mode   &  width(MeV)\\\hline
  $Z_c(3900)\rightarrow J/\psi\pi$  &   3.67 &  $Z_c(4030)\rightarrow J/\psi\pi$  &  17.85  \\
  $Z_c(3900)\rightarrow \psi(2S)\pi$  &   8.24 &  $Z_c(4030)\rightarrow \psi(2S)\pi$  &  0.30    \\
  $Z_c(3900)\rightarrow \rho\eta_c$  &  0.45 & $Z_c(4030)\rightarrow \rho\eta_c$
  &0.30
  \\ $Z_c(3900)\rightarrow DD^*$  &  0.024& $Z_c(4030)\rightarrow DD^*$  &
 0.23 \\
  - & - & $Z_c(4030)\rightarrow D^*D^*$  &
0.52
\end{tabular}
\end{ruledtabular}
\end{table}

%\begin{table}[!h]
%\caption{The decay widths of some modes ($\Lambda=0.88$GeV,
%$\beta=0.2$GeV$^{-1}$).}\label{Tab:t3}
%\begin{ruledtabular}
%\begin{tabular}{cc|cc}
 % decay mode   &  width(MeV)&  decay mode   &  width(MeV)\\\hline
%  $Z_c(3900)\rightarrow J/\psi\pi$  &   0.54 &  $Z_c(4030)\rightarrow J/\psi\pi$  &  1.97   \\
 %  $Z_c(3900)\rightarrow \psi(2S)\pi$  &   2.37&  $Z_c(4030)\rightarrow \psi(2S)\pi$  &   3.67 $\times 10^{-4}$   \\
%   $Z_c(3900)\rightarrow \rho\eta_c$  &  0.045 & $Z_c(4030)\rightarrow \rho\eta_c$
%  &0.030\\
%  $Z_c(3900)\rightarrow DD^*$  &  2.47$\times 10^{-4}$ & $Z_c(4030)\rightarrow DD^*$  &
% 0.14 \\
%  - & - & $Z_c(4030)\rightarrow D^*D^*$  &
% 0.078
%\end{tabular}
%\end{ruledtabular}
%\end{table}

\begin{table}[!h]
\caption{The decay widths of some modes ($\Lambda=1.1$ GeV,
$\beta=0.631$ GeV$^{-1}$).}\label{Tab:t4}
\begin{ruledtabular}
\begin{tabular}{cc|cc}
  decay mode   &  width(MeV)&  decay mode   &  width(MeV)\\\hline
  $Z_c(3900)\rightarrow J/\psi\pi$  &   6.44 &  $Z_c(4030)\rightarrow J/\psi\pi$  &   35.08   \\
   $Z_c(3900)\rightarrow \psi(2S)\pi$  &  12.00&  $Z_c(4030)\rightarrow \psi(2S)\pi$  &   0.66  \\
   $Z_c(3900)\rightarrow \rho\eta_c$  &  0.88 & $Z_c(4030)\rightarrow \rho\eta_c$  &
  0.53\\
  $Z_c(3900)\rightarrow DD^*$  &  0.055 & $Z_c(4030)\rightarrow DD^*$  &
  0.62\\
  - & - & $Z_c(4030)\rightarrow D^*D^*$  &
0.96
\end{tabular}
\end{ruledtabular}
\end{table}

\begin{figure}
\begin{center}
\begin{tabular}{ccc}
\scalebox{0.8}{\includegraphics{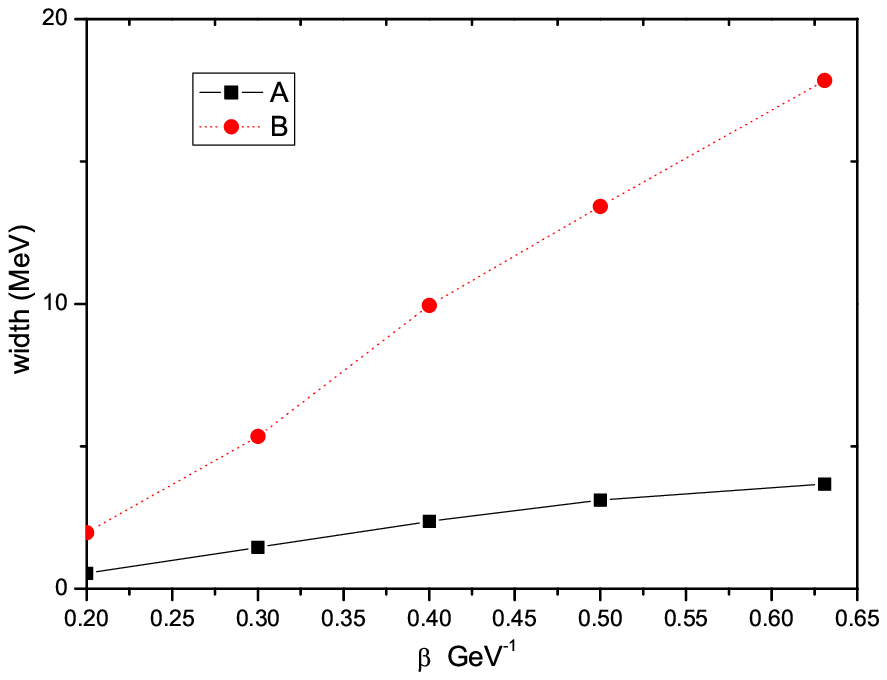}}
\end{tabular}
\end{center}
\caption{the dependence of $\Gamma(Z_c(3900)\rightarrow J/\psi
\pi$) (A) and $\Gamma(Z_c(4030)\rightarrow J/\psi \pi$)  (B) on
$\beta$.}\label{Fig6}
\end{figure}

One can find if  $Z_c(3900)$ is a molecular state of $D\bar D^*$
the partial width of $Z_c(3900)\rightarrow D\bar D^*$ is very
small. For $Z_c(4030)$  both the partial widths
$\Gamma(Z_c(4030)\rightarrow D^*\bar D^*)$
$\Gamma(Z_c(4030)\rightarrow D\bar D^*)$ are small, but
sufficiently sizable to be  measured.

In our calculation, we notice that the model parameters $\Lambda$
and $\beta$ can affect the numerical results sensitively. Since
the $\beta$ is a model parameter which is closely related to the
behavior of the wavefuctions of the concerned hadrons, we
illustrate the dependence of $Z_c(3900)\rightarrow J/\psi\pi$ and
$Z_c(4030)\rightarrow J/\psi\pi$ on $\beta$ in Fig.\ref{Fig6}.
Line A and B in Fig.\ref{Fig6} correspond to $Z_c(3900)$ and
$Z_c(4030)$ respectively. In Ref.\cite{Dong:2013iqa} the estimated
$\Gamma(Z_c(3900)\rightarrow J/\psi\pi)$ is about 1 MeV which is
smaller than our results.
%If one wants to get
%this order of magnitude of $\Gamma(Z_c(3900)\rightarrow
%/\psi\pi)$ the parameter $\beta$ or $\Lambda$ should be lower
%than the numbers in  this paper. In Tab.\ref{Tab:t3} we present
%the theoretical estimates with $\beta=0.2$GeV$^{-1}$ and
%$\Lambda=0.88$GeV which indicate a small $\beta$ maybe consist
%with data.

\section{discussions and a tentative conclusion}

In this work we calculate the decay rates of $Z_c(3900)$ to
$J\psi\pi$, $\psi(2s)\pi$, $\eta_c\rho$ and $D^*\bar D^*$ in the
LFM by assuming that $Z_c(3900)$ is a molecular state of
$\frac{1}{\sqrt 2}(D\bar D^*+D^*\bar D)$. The numerical results
are shown in tables II and III where we vary the model parameters
$\Lambda$ to check the parameter dependence. Then in the same
framework, we evaluate the decay rates of a postulated $Z_c(4030)$
which was predicted by Voloshin\cite{Voloshin:2013dpa} according
to the mass gap between $Z_b(10610)$ and $Z_b(10650)$. It is
tempted to consider $Z_c(4030)$ as a molecular state of $D^*\bar
D^*$. Very recently, the BES collaboration reports that two
resonances $Z_c(4020)$ and $Z_c(4025)$ were observed \cite{Yuan}.
The resonance $Z_c(4020)$ is observed in the channel of $e^+e^-\to
\pi Z_c (4020)\to \pi^+\pi^-h_c(1P)$ at CM energies of 4.26/4.36
GeV and the fit results are $M(4020)=(4021.8\pm 1.0\pm 2.5)$ MeV
and $\Gamma(4020)=(5.7\pm 3.4\pm 1.1)$ MeV, whereas $Z_c(2025)$ is
observed in the channel $e^+e^-\to \pi^-D^*\bar D^*$ at the CM
energy of 4.26 GeV with the fit results of $N(4025)=(4026.3\pm
2.6\pm 3.7)$ MeV, $\Gamma(4025)=(24.8\pm 5.7\pm 7.7)$ MeV. Both of
them are slightly lower than 4030 MeV. Since they are apart by
only 2$\sigma$, one still may ask if they are the same resonance
and the difference is due to experimental errors.The results of
this work may help to clarify if they are indeed  one resonance.

Our numerical results listed in tables  II and III show that the
decay rate of $Z_c(3900)\to J/\psi \pi$, which is the channel of
first observing $Z_c(3900)$, is about a few MeV. But the decay
$Z_c(3900)\to D\bar D^*( {\rm or} D^*\bar D)$ is much smaller than
that for the $J/\psi$ final state. It is also noted that the rate
of $Z_c(3900)\to \psi(2s)\pi$ is almost twice larger than that for
$Z_c(3900)\to J/\psi \pi$. Even though the final state phase space
of $\psi(2s)\pi$ is smaller than that for $J/\psi\pi$, the form
factor for  $Z_c\to \psi(2s)\pi$ is larger to result in the
enhanced rate.

For the numerical computations we take 4030 MeV as the  mass of
$Z_c(4030)$ , but of course it is easy to adjust it into 4020 or
4025 MeV. Since the two resonances are waiting for further
confirmation, we at present employ 4030 MeV as input, and the
numerical results would be of significance even though not
accurate. Our numerical results show that the tendency of the
predicted decay rates for $Z_c(4030)$ are similar to that for
$Z_c(3900)$.

The predicted rates are somehow sensitive to the model parameters,
for example in tables II and III, we only change the value of
$\Lambda$ from 0.88 GeV to 1.1 GeV which were determined by
fitting data of different experiments, the rates are almost
doubled. Fig.6 shows dependence of $\Gamma(Z_c(3900)\to
J/\psi\pi)$ and $\Gamma(Z_c(4030)\to J/\psi\pi)$ on the
$\beta-$value. Since the values of $\Lambda$ and $\beta$ are
obtained by fitting data of experiments, we can only adjust those
parameters within small ranges, so that the predictions with the
model cannot be drastically changed. Namely, even though the
predicted values may vary by a factor of two or even larger, the
degree of magnitude remains the same.

Because the theoretical predictions are not fully accordant
with the available data, although the data so far are not very accurate, we prefer to draw a
tentative conclusion that the observed $Z_c(3900)$ and newly observed $Z_c(4020)$ and/or $Z_c(4025)$ are not molecular states
of $D\bar D^*$ and $D^*\bar D^*$, but may be tetraquarks  or mixtures of molecular
states and tetraquarks. It is difficult to evaluate the decay rates
of tetraquarks because there the non-perturbative QCD effects would dominate. In our later
work, we will try to do it in terms of some reasonable models. Definitely further more
accurate measurements on the decays of such exotic states: $Z_b$, $Z_b'$, $Z_c$ and $Z_c'$ are very badly needed.

\section*{Acknowledgement}

We thank Dr. C.X. Yu and Dr. Y.P. Guo for introducing some details about the measurements to us and drawing our attention to Dr. Yuan's talk at the lepton-photon conference.
This work is supported by the National Natural Science Foundation
of China (NNSFC) under the contract No. 11075079 and No. 11005079;
the Special Grant for the Ph.D. program of Ministry of Eduction of
P.R. China No. 20100032120065.

\appendix

\section{}

\appendix
\section{the vertex function of molecular state}

Supposing $Z_c(3900)$ and $Z_c(4030)$ are  molecular states which
consists of $D$ and $\bar {D^*}$ and $D^{*}$ and $\bar {D^*}$
respectively. If the orbital angular momentum between the two
components is zero, i.e. $l=0$, the total spin $S$ should be 0, 1
and 2 and total angular momentum $J$ also is 0, 1 and 2.

Similar to our previous works on baryons \cite{Ke:2007tg}, we
construct the vertex function of molecular  in the same model. The
wavefunction of a molecualr with total spin $J$ and momentum $P$
is
\begin{eqnarray}\label{eq:lfbaryon}
 |X(P,J,J_z)\rangle&=&\int\{d^3\tilde p_1\}\{d^3\tilde p_2\} \,
  2(2\pi)^3\delta^3(\tilde{P}-\tilde{p_1}-\tilde{p_2}) \nonumber\\
 &&\times\sum_{\lambda_1}\Psi^{SS_z}(\tilde{p}_1,\tilde{p}_2,\lambda_1,\lambda_2)
  \mathcal{F}\left|\right.
  D^{(*)}(p_1,\lambda_1) \bar D^*(p_2,\lambda_2)\ra,
\end{eqnarray}
with
\begin{eqnarray*}
 \Psi^{SS_z}(\tilde{p}_1,\tilde{p}_2,\lambda_1,\lambda_2)=&&
  \left\la\lambda_1\left|\mathcal{R}^{\dagger}_M(x_1,k_{1\perp},m_1)
   \right|s_1\right\ra  \left\la\lambda_2\left|\mathcal{R}^{\dagger}_M(x_2,k_{2\perp},m_2)
   \right|s_2\right\ra
  \nonumber\\ &&\left\la
1s_1 ;1 s_2\left|S S_z\right\ra \la S S_z ;0 0\left|J J_z\right\ra
 \varphi(x,k_{\perp})\right.,
\end{eqnarray*}
where $\la 1s_1 ;1 s_2|S S_z\ra \la S S_z ;0 0|J J_z\ra$ is the
C-G coefficients and $s_1,s_2$ are the spin projections of the
constituents. These C-G coefficients can be rewrote as
\begin{eqnarray}
&&\la  1s_1;0 0|S S_z\ra \la S S_z ;0 0|1 J_z\ra=A_{10}
 \epsilon_{1}(s_1)\cdot
 \epsilon(J_z)\nonumber\\&&\la  0 0;1s_2|S S_z\ra \la S S_z ;0 0|1
 J_z\ra=A_{01}
 \epsilon_{2}(s_2)\cdot
 \epsilon(J_z)\nonumber\\&&\la 1s_1 ;1 s_2|S S_z\ra \la S S_z ;0 0|0 0\ra=A_0
 \epsilon_{1}(s_1)\cdot
 \epsilon_{2}(s_2)\nonumber\\&&\la 1s_1 ;1 s_2|S S_z\ra \la S S_z ;0 0|1 J_z\ra=A_1
\varepsilon^{\mu\nu\alpha\beta} \epsilon_{1\mu}(s_1)
 \epsilon_{2\nu}(s_2)\epsilon_{\alpha}(J_z)P_\beta
 \nonumber\\&&\la 1s_1 ;1 s_2|S S_z\ra \la S S_z ;0 0|2
 J_z\ra=A_2
\epsilon_{1\mu}(s_1)
 \epsilon_{2\nu}(s_2)\epsilon^{\mu\nu}(J_z)
\end{eqnarray}
with
\begin{eqnarray*}
&&A_{01}=\frac{\sqrt{3}m_1}{\sqrt{e_1^2+2m_1^2}},\nonumber\\&&
A_{10}=\frac{\sqrt{3}m_2}{\sqrt{e_2^2+2m_2^2}},\nonumber\\&&A_0=\frac{2
{m_1} {m_2}}{\sqrt{{M_0'}^4-2 {M_0'}^2
({m_1}^2+{m_2}^2)+{m_1}^4+10 {m_1}^2 {m_2}^2+{m_2}^4}},
\nonumber\\&& A_1=\frac{2\sqrt{3} {m_1} {m_2}}{\sqrt{{M'}^2 [4
{e_1}^2 {m_2}^2-4 {e_1} {e_2} (-{M_0'}^2+{m_1}^2+{m_2}^2)+4
{e_2}^2
   {m_1}^2+10 {m_1}^2 {m_2}^2-C_A]}}
, \nonumber\\&&A_2=\frac{\sqrt{120} {m_1} {m_2}}{\sqrt{4 {e_1}^2
(4 {e_2}^2+7 {m_2}^2)+4 {e_1} {e_2} (-{M_0'}^2+{m_1}^2+{m_2}^2)+28
{e_2}^2
   {m_1}^2+54 {m_1}^2
   {m_2}^2+C_A}}, \nonumber\\&&C_A={M_0'}^4-2 {M_0'}^2
   ({m_1}^2+{m_2}^2)+{m_1}^4+{m_2}^4.
\end{eqnarray*}

 A Melosh transformation brings the the matrix
elements from the spin-projection-on-fixed-axes representation
into the helicity representation and  is explicitly written as
$$\left\la\lambda_2\left|\mathcal{R}^{\dagger}_M(x_2,k_{2\perp},m_2)
   \right|s_2\right\ra=\xi^*(\lambda_2,m_2)\cdot\xi(s_2,m_2),$$ and $$\left\la\lambda_1\left|\mathcal{R}^{\dagger}_M(x_1,k_{1\perp},m_1)
   \right|s_1\right\ra=\xi^*(\lambda_1,m_1)\cdot\xi(s_1,m_1).$$

Following Refs. \cite{Jaus,Cheng:2003sm}, the Melosh transformed
matrix can be expressed as
\begin{eqnarray}
 && \la\lambda_1|\mathcal{R}^{\dagger}_M(x_1,k_{1\perp},m_1)
  |s_1\ra  \la\lambda_2|\mathcal{R}^{\dagger}_M(x_2,k_{2\perp},m_2)
   |s_2\ra
  \nonumber\\ &&\la
1s_1 ;1 s_2|S S_z\ra \la S S_z ;0 0|J J_z\ra\nonumber\\&&=A_1
\varepsilon^{\mu\nu\alpha\beta} \epsilon_{1\mu}(\lambda_1)
 \epsilon_{2\nu}(\lambda_2)\epsilon_{\alpha}(J_z)P_\beta,
\end{eqnarray}
so the wavefunction of $1^+$ molecular state of $D^{(*)}\bar D^*$
\begin{eqnarray}
 \Psi^{SS_z}(\tilde{p}_1,\tilde{p}_2,\lambda_1,\lambda_2)&&=
 A_1 \varphi(x,k_{\perp})
\varepsilon^{\mu\nu\alpha\beta} \epsilon_{1\mu}(\lambda_1)
 \epsilon_{2\nu}(\lambda_2)\epsilon_{\alpha}(J_z)P_\beta\nonumber\\&&
=h_{A_1}' \varepsilon^{\mu\nu\alpha\beta}
\epsilon_{1\mu}(\lambda_1)
 \epsilon_{2\nu}(\lambda_2)\epsilon_{\alpha}(J_z)P_\beta,
\end{eqnarray}

with
$\varphi=4(\frac{\pi}{\beta^2})^{3/4}\frac{e_1e_2}{x_1x_2M_0}{\rm
exp}(\frac{-\mathbf{k}^2}{2\beta^2})$.

Similarly   the wavefunction of $1^+$ molecular state of $D\bar
D^*$
\begin{eqnarray}
 \Psi^{SS_z}(\tilde{p}_1,\tilde{p}_2,\lambda_1,\lambda_2)&&=
 A\varphi(x,k_{\perp})
\epsilon_{1\mu}(\lambda_1)
 \cdot\epsilon_{\alpha}(J_z)\nonumber\\&&
=h_{A_1}' \epsilon_{1\mu}(\lambda_1)
 \cdot\epsilon_{\alpha}(J_z).
\end{eqnarray}

 and  normalization of  the
state is $X(P,J,J_z)\rangle$ , \beq\label{A12}
 \la X(P',J',J_z')|X(P,J,J_z)\ra=2(2\pi)^3P^+
  \delta^3(\tilde{P}'-\tilde{P})\delta_{J'J}\delta_{J'_zJ_z}.
 \eeq

{ All other notations can be found in Ref}.\cite{Ke:2007tg}.
\section{the effective vertices}
 the effective vertices  can be found in
\cite{Haglin:1999xs,Oh:2000qr,Lin:1999ad,Deandrea:2003pv,Meng:2007cx},
\begin{eqnarray}\label{lagrangian_piDD}
 &&\mathcal L_{\pi DD^*}=ig_{_{\pi DD^*}}(D^{*\mu}\partial_\mu\pi\bar D - \partial^\mu D  \pi\bar D^{*}_\mu+h.c.)
 ,\\
 &&\mathcal L_{\pi D^*D^*}=-g_{_{\pi D^*D^*}}\varepsilon^{\mu\nu\alpha\beta}\partial^\mu\bar D^*_\nu\pi\partial_\alpha
 D^{*\beta},\\&&\mathcal L_{\psi DD}=ig_{_{\psi DD}}\psi_\mu(\partial^\mu D\bar D - D\partial^\mu  \bar D)
 ,\\
 &&\mathcal L_{\psi DD^*}=-g_{_{\psi DD^*}}\varepsilon^{\mu\nu\alpha\beta}\partial^\mu\psi_\nu
 [{\partial}_\beta D^*_\alpha \bar D+D{  \partial}_\beta \bar D^*_\alpha]
 \\&&\mathcal L_{\psi D^*D^*}=ig_{_{\psi D^*D^*}}
 [-\psi^\mu D^{*\nu}(\overrightarrow{\partial}-\overleftarrow{\partial})_\mu D_\nu^{*\dagger}+
 \psi^\mu D^{*\nu}{\partial}_\nu D_\mu^{*\dagger}-\psi^\mu {\partial}_\nu D^{*\mu} D_\nu^{*\dagger}],\\&&
\mathcal L_{\sigma DD}=-g_{_{\sigma DD}}\sigma D\bar D
\\&&\mathcal L_{\sigma D^*D^*}=g_{_{\sigma D^*D^*}}\sigma D^* \cdot\bar D^*
 %\mathcal L_{\rho DD^*}=-g_{\rho
%DD^*}\varepsilon^{\mu\nu\alpha\beta}\partial^\mu\rho_\nu
% [{\partial}_\beta D^*_\alpha \bar D+D{ \partial}_\beta  {\bar D}*_\alpha]
% \\&&\mathcal L_{\rho D^*D^*}=ig_{\rho D^*D^*}
% [-\rho^\mu D^{*\nu}(\overrightarrow{\partial}-\overleftarrow{\partial})_\mu D_\nu^{*\dagger}+
% \rho^\mu D^{*\nu}{\partial}_\nu D_\mu^{*\dagger}-\psi^\mu {\partial}_\nu D^{*\mu} D_\nu^{*\dagger}]
\end{eqnarray}
The  effective vertices $\eta_c DD^*$ and  $\eta_c D^*D^*$ are
similar to those in Eq. (B1) and  Eq. (B2) and the effective
vertices $\rho DD$,  $\rho DD^*$ and $\rho D^*D^*$ can be obtained
by replacing the $\psi$ by $\rho$ in Eq. (B3) and Eq. (B4).


\begin{thebibliography}{99}
%\cite{Aubert:2003fg}
%\cite{Ablikim:2013mio}
\bibitem{Ablikim:2013mio}
  M.~Ablikim {\it et al.}  [ BESIII Collaboration],
  %``Observation of a charged charmoniumlike structure in e+e- to pi+pi-J/psi at \sqrt{s}=4.26 GeV,''
   arXiv:1303.5949 [hep-ex].  %%CITATION = ARXIV:1303.5949;%%  %26 citations counted in INSPIRE as of 06 Jun 2013
%\cite{Chen:2013coa}


%\cite{Liu:2013dau}
\bibitem{Liu:2013dau}
  Z.~Q.~Liu {\it et al.}  [Belle Collaboration],
  %``Study of $e^+ e^- \to \pi^+ \pi^- J/\psi$ and Observation of a Charged Charmonium-like State at Belle,''
  arXiv:1304.0121 [hep-ex].  %%CITATION = ARXIV:1304.0121;%%  %18 citations counted in INSPIRE as of 06 Jun 2013
%\cite{Xiao:2013iha}
\bibitem{Xiao:2013iha}
  T.~Xiao, S.~Dobbs, A.~Tomaradze and K.~K.~Seth,
  %``Observation of the Charged Hadron Zc(3900) at sqrt(s)=4170 MeV,''
   arXiv:1304.3036 [hep-ex].  %%CITATION = ARXIV:1304.3036;%%  %11 citations counted in INSPIRE as of 06 Jun 2013
%\cite{Wang:2013cya}

%\cite{Choi:2003ue}
\bibitem{Choi:2003ue}
  S.~K.~Choi {\it et al.}  [Belle Collaboration],
  %``Observation of a narrow charmonium - like state in exclusive B+- ---> K+-
  %pi+ pi- J / psi decays,''
  Phys.\ Rev.\ Lett.\  {\bf 91}, 262001 (2003)
  [arXiv:hep-ex/0309032].
  %%CITATION = PRLTA,91,262001;%%

%\cite{Collaboration:2011gj}
\bibitem{Collaboration:2011gj}
  B.~Collaboration,
  %``Observation of two charged bottomonium-like resonances,''
  arXiv:1105.4583 [hep-ex].
  %%CITATION = ARXIV:1105.4583;%%

\bibitem{Chen:2013coa}
  D.~-Y.~Chen, X.~Liu and T.~Matsuki,
  %``Reproducing $Z_c(3900)$ through the ISPE mechanism,''
  arXiv:1304.5845 [hep-ph].  %%CITATION = ARXIV:1304.5845;%%  %1 citations counted in INSPIRE as of 06 Jun 2013

\bibitem{Wang:2013cya}
  Q.~Wang, C.~Hanhart and Q.~Zhao,
  %``Decoding the riddle of Y(4260) and $Z_c(3900)$,''
  arXiv:1303.6355 [hep-ph].  %%CITATION = ARXIV:1303.6355;%%  %14 citations counted in INSPIRE as of 06 Jun 2013

%\cite{Wilbring:2013cha}
\bibitem{Wilbring:2013cha}
  E.~Wilbring, H.~-W.~Hammer and U.~-G.~Mei?ner,
  %``Electromagnetic Structure of the Z_c(3900),''
  arXiv:1304.2882 [hep-ph].  %%CITATION = ARXIV:1304.2882;%%  %7 citations counted in INSPIRE as of 06 Jun 2013




%\cite{Cui:2013yva}
\bibitem{Cui:2013yva}
  C.~-Y.~Cui, Y.~-L.~Liu, W.~-B.~Chen and M.~-Q.~Huang,
  %``Could $Z_{c}(3900)$ be a $I^{G}J^{P}=1^{+}1^{+}$ $D^{*}\bar{D}$ molecular state?,''
  arXiv:1304.1850 [hep-ph].  %%CITATION = ARXIV:1304.1850;%%  %7 citations counted in INSPIRE as of 06 Jun 2013
%\cite{Zhang:2013aoa}
\bibitem{Zhang:2013aoa}
  J.~-R.~Zhang,
  %``Improved QCD sum rule study of $Z_{c}(3900)$ as a $\bar{D}D^{*}$ molecular state,''
   arXiv:1304.5748 [hep-ph].  %%CITATION = ARXIV:1304.5748;%%  %3 citations counted in INSPIRE as of 06 Jun 2013

%\cite{Voloshin:2013dpa}
\bibitem{Voloshin:2013dpa}
  M.~B.~Voloshin,
  %``Z_c(3900) - what is inside?,''
  arXiv:1304.0380 [hep-ph],  Phys.Rev.{\bf D87}, 091501(R), (2013).






\bibitem{Jaus} W. Jaus, Phys. Rev.  D {\bf 41}, 3394 (1990);
  D {\bf 44}, 2851 (1991);  W.~Jaus,
  %``Covariant analysis of the light-front quark model,''
  Phys.\ Rev.\  D {\bf 60}, 054026 (1999).
  %%CITATION = PHRVA,D60,054026;%%

%\cite{Ji:1992yf}
\bibitem{Ji:1992yf}
  C.~R.~Ji, P.~L.~Chung and S.~R.~Cotanch,
  %``Light Cone Quark Model Axial Vector Meson Wave Function,''
  Phys.\ Rev.\  D {\bf 45}, 4214 (1992).
  %%CITATION = PHRVA,D45,4214;%%
%\cite{Cheng:1996if}
  %\cite{Cheng:2004cc}
\bibitem{Cheng:2004cc}
  H.~-Y.~Cheng, C.~-K.~Chua and C.~-W.~Hwang,
  %``Light front approach for heavy pentaquark transitions,''
   Phys.\ Rev.\ D {\bf 70}, 034007 (2004)  [hep-ph/0403232].  %%CITATION = HEP-PH/0403232;%%  %22 citations counted in INSPIRE as of 06 Jun 2013


%\cite{Ke:2007tg}
\bibitem{Ke:2007tg}
  H.~W.~Ke, X.~Q.~Li and Z.~T.~Wei,
  %``Diquarks and \Lambda_{b}\to\Lambda_c weak decays,''
  Phys.\ Rev.\  D {\bf 77}, 014020 (2008)
  [arXiv:0710.1927 [hep-ph]];
  %%CITATION = PHRVA,D77,014020;%%
%\bibitem{Wei:2009np}
  Z.~T.~Wei, H.~W.~Ke and X.~Q.~Li,
  %``Evaluating decay Rates and Asymmetries of $\Lambda_b$ into Light Baryons in
  %LFQM,''
  Phys.\ Rev.\  D {\bf 80}, 094016 (2009)
  [arXiv:0909.0100 [hep-ph]];
  %%CITATION = PHRVA,D80,094016;%%
%\cite{Ke:2009ed}
%\cite{Ke:2012wa}
%\bibitem{Ke:2012wa}
  H.~-W.~Ke, X.~-H.~Yuan, X.~-Q.~Li, Z.~-T.~Wei and Y.~-X.~Zhang,
  %``$\Sigma_{b}\to\Sigma_c$ and $\Omega_b\to\Omega_c$ weak decays in the light-front quark model,''
   Phys.\ Rev.\ D {\bf 86}, 114005 (2012)  [arXiv:1207.3477 [hep-ph]].  %%CITATION = ARXIV:1207.3477;%%  %1 citations counted in INSPIRE as of 06 Jun 2013


\bibitem{Ke:2009ed}
  H.~W.~Ke, X.~Q.~Li and Z.~T.~Wei,
  %``Whether new data on $D_s\to f_0(980) e^+ \nu_e$ can be understood if
  %$f_0(980)$ consists of only the conventional $q\bar{q}$ structure,''
  Phys.\ Rev.\  D {\bf 80}, 074030 (2009)
  [arXiv:0907.5465 [hep-ph]];
%
%\cite{Ke:2009mn}
%\bibitem{Ke:2009mn}
  H.~W.~Ke, X.~Q.~Li and Z.~T.~Wei,
  %``Determining the $\eta-\eta'$ mixing by the newly measured
  %$BR(D(D_s)\to\eta(\eta')+\bar l+\nu_l$,''
  Eur.\ Phys.\ J.\  C {\bf 69}, 133 (2010)
  [arXiv:0912.4094 [hep-ph]];
  %%CITATION = EPHJA,C69,133;%%
%\cite{Ke:2011fj}
%\bibitem{Ke:2011fj}
  H.~W.~Ke, X.~H.~Yuan and X.~Q.~Li,
  %``Fraction of the gluonium component in $\eta'$ and $\eta$,''
Int. J. Mod. Phys. A {\bf 26}, 4731 (2011),
  arXiv:1101.3407 [hep-ph];
    %%CITATION = ARXIV:1101.3407;%%
%\cite{Ke:2011mu}
%\bibitem{Ke:2011mu}
  H.~W.~Ke and X.~Q.~Li,
  %``Vertex functions for d-wave mesons in the light-front approach,''
  Eur.\ Phys.\ J.\  C {\bf 71}, 1776 (2011)
  [arXiv:1104.3996 [hep-ph]].
  %%CITATION = EPHJA,C71,1776;%%




\bibitem{Cheng:1996if}
  H.~Y.~Cheng, C.~Y.~Cheung and C.~W.~Hwang,
  %``Mesonic form factors and the Isgur-Wise function on the light-front,''
  Phys.\ Rev.\  D {\bf 55}, 1559 (1997)
  [arXiv:hep-ph/9607332].
  %%CITATION = PHRVA,D55,1559;%%

%\cite{Li:2010bb}
\bibitem{Li:2010bb}
  G.~Li, F.~l.~Shao and W.~Wang,
  %``$B_s\to D_s(3040)$ form factors and $B_s$ decays into $D_s(3040)$,''
  Phys.\ Rev.\  D {\bf 82}, 094031 (2010)
  [arXiv:1008.3696 [hep-ph]].
  %%CITATION = PHRVA,D82,094031;%%
%\cite{Colangelo:2010bg}
%\bibitem{Colangelo:2010bg}
%  P.~Colangelo, F.~De Fazio and W.~Wang,
  %``$B_s\to f_0(980)$ form factors and $B_s$ decays into $f_0(980)$,''
%  Phys.\ Rev.\  D {\bf 81}, 074001 (2010)
%  [arXiv:1002.2880 [hep-ph]].
  %%CITATION = PHRVA,D81,074001;%%

%\cite{Cheng:2003sm}
\bibitem{Cheng:2003sm}
  H.~Y.~Cheng, C.~K.~Chua and C.~W.~Hwang,
  %``Covariant light-front approach for s-wave and p-wave mesons: Its
  %application to decay constants and form factors,''
  Phys.\ Rev.\  D {\bf 69}, 074025 (2004).
  %%CITATION = PHRVA,D69,074025;%%
%\cite{Zweber:2007zz}


%\cite{Hwang:2006cua}
\bibitem{Hwang:2006cua}
  C.~W.~Hwang and Z.~T.~Wei,
  %``Covariant light-front approach for heavy quarkonium: decay constants, P \to
  %\gamma \gamma and V \to P \gamma,''
  J.\ Phys.\ G {\bf 34}, 687 (2007);
  %%CITATION = JPHGB,G34,687;%%
%
 %%\cite{Lu:2007sg}
  %\bibitem{Lu:2007sg}
  C.~D.~Lu, W.~Wang and Z.~T.~Wei,
  %``Heavy-to-light form factors on the light cone,''
  Phys.\ Rev.\  D {\bf 76}, 014013 (2007)
  [arXiv:hep-ph/0701265].
  %%CITATION = PHRVA,D76,014013;%%


%\cite{Choi:2007se}
\bibitem{Choi:2007se}
  H.~M.~Choi,
  %``Decay constants and radiative decays of heavy mesons in light-front   quark
  %model,''
  Phys.\ Rev.\  D {\bf 75}, 073016 (2007)
  [arXiv:hep-ph/0701263];
  %%CITATION = PHRVA,D75,073016;%%

%\cite{Ke:2013zs}
\bibitem{Ke:2013zs}
  H.~-W.~Ke, X.~-Q.~Li and Y.~-L.~Shi,
  %``The radiative decays of $0^{++}$ and $1^{+-}$ heavy mesons,''
 Phys.\ Rev.\  D {\bf 87}, 054022 (2013)
  arXiv:1301.4014 [hep-ph]; %%CITATION = ARXIV:1301.4014;%%

%\bibitem{Ke:2010x}
  H.~W.~Ke, X.~Q.~Li, Z.~T.~Wei and X.~Liu,
  %``Re-Study on the wave functions of $\Upsilon(nS)$ states in LFQM and the
  %radiative decays of $\Upsilon(nS)\to \eta_b+\gamma$,''
  Phys.\ Rev.\  D {\bf 82}, 034023 (2010)
  [arXiv:1006.1091 [hep-ph]].
  %%CITATION = PHRVA,D80,015022;%%



%\cite{Haglin:1999xs}
\bibitem{Haglin:1999xs}
  K.~L.~Haglin,
  %``Charmonium dissociation in hadronic matter,''
  Phys.\ Rev.\ C {\bf 61} (2000) 031902.%  [nucl-th/9907034].  %%CITATION = NUCL-TH/9907034;%%

%\cite{Oh:2000qr}
\bibitem{Oh:2000qr}
  Y.~-S.~Oh, T.~Song and S.~H.~Lee,
  %``J / psi absorption by pi and rho mesons in meson exchange model with anomalous parity interactions,''
  Phys.\ Rev.\ C {\bf 63}, 034901 (2001)
  [nucl-th/0010064].
  %%CITATION = NUCL-TH/0010064;%%

\bibitem{Lin:1999ad}
  Z.~-W.~Lin and C.~M.~Ko,
  %``A Model for J / psi absorption in hadronic matter,''
  Phys.\ Rev.\ C {\bf 62}, 034903 (2000).%  [nucl-th/9912046].  %%CITATION = NUCL-TH/9912046;%%

\bibitem{Deandrea:2003pv}
 A.~Deandrea, G.~Nardulli and A.~D.~Polosa,
 %``J / psi couplings to charmed resonances and to pi,''
 Phys.\ Rev.\ D {\bf 68}, 034002 (2003)[hep-ph/0302273].
 %%CITATION = HEP-PH/0302273;%%

%\cite{Meng:2007cx}
\bibitem{Meng:2007cx}
  C.~Meng and K.~-T.~Chao,
  %``Decays of the $X(3872) $ and chi(c1) (2P) charmonium,''
  Phys.\ Rev.\ D {\bf 75}, 114002 (2007)
  [hep-ph/0703205].
  %%CITATION = HEP-PH/0703205;%%
  %49 citations counted in INSPIRE as of 27 May 2013



%\cite{Yuan:2012zw}
\bibitem{Yuan:2012zw}
  X.~-H.~Yuan, H.~-W.~Ke, X.~Liu and X.~-Q.~Li,
  %``Can FSI explain the observed $B(B_s\rightarrow K^+K^-){>}B(B_s\rightarrow \pi^+K^-)$ anomaly?,''
   Phys.\ Rev.\ D {\bf 87}, 014019 (2013)  [arXiv:1210.3686 [hep-ph]].  %%CITATION = ARXIV:1210.3686;%%

%\cite{Beringer:1900zz}


%\cite{Guo:2007mm}
\bibitem{Guo:2007mm}
  X.~H.~Guo and X.~H.~Wu,
  %``Studying the scalar bound states of K anti-K system in Bethe-Salpeter
  %formalism,''
  Phys.\ Rev.\  D {\bf 76} (2007) 056004
  [arXiv:0704.3105 [hep-ph]];
  %%CITATION = PHRVA,D76,056004;%%
%\cite{Ke:2012gm}
%\bibitem{Ke:2012gm}
  H.~-W.~Ke, X.~-Q.~Li, Y.~-L.~Shi, G.~-L.~Wang and X.~-H.~Yuan,
  %``Is $Z_b(10610)$ a Molecular State?,''
   JHEP {\bf 1204}, 056 (2012)  [arXiv:1202.2178 [hep-ph]].  %%CITATION = ARXIV:1202.2178;%%  %3 citations counted in INSPIRE as of 06 Jun 2013

\bibitem{PDG12}
  J.~Beringer {\it et al.}  [Particle Data Group Collaboration],
  %``Review of Particle Physics (RPP),''
  Phys.\ Rev.\ D {\bf 86}, 010001 (2012).  %%CITATION = PHRVA,D86,010001;%%


  %\cite{Lee:2009hy}
\bibitem{Lee:2009hy}
  I.~W.~Lee, A.~Faessler, T.~Gutsche and V.~E.~Lyubovitskij,
  %``X(3872) as a molecular DD* state in a potential model,''
  Phys.\ Rev.\  D {\bf 80}, 094005 (2009)
  [arXiv:0910.1009 [hep-ph]].
  %%CITATION = PHRVA,D80,094005;%%

%\cite{Ke:2011jf}
\bibitem{Ke:2011jf}
  H.~W.~Ke and X.~Q.~Li,
  %``What do the radiative decays of X(3872) tell us,''
  Phys.\ Rev.\  D {\bf 84}, 114026 (2011)
  [arXiv:1107.0443 [hep-ph]];
  %%CITATION = PHRVA,D84,114026;%%

%\cite{Dong:2013iqa}
\bibitem{Dong:2013iqa}
  Y.~Dong, A.~Faessler, T.~Gutsche and V.~E.~Lyubovitskij,
  %``Strong decays of molecular states Zc(+) and Zc'(+),''
   arXiv:1306.0824 [hep-ph].  %%CITATION = ARXIV:1306.0824;%%

\bibitem{Yuan} C.Z. Yuan, Talk presented at The XXVI International Symposium
on Lepton-Photon interactions at High Energies, San Francisco, USA, June 25, 2013.


\end{thebibliography}
\end{document}